\documentclass[twocolumn,trackchanges]{aastex631}

\submitjournal{ApJ}

\usepackage{amsmath,amssymb,mathtools}
\usepackage[textsize=small]{todonotes}
\usepackage{savesym}
\usepackage{subfigure}
\savesymbol{tablenum}
\usepackage{siunitx}
\restoresymbol{SIX}{tablenum}
\usepackage{subfigure}
\usepackage{enumitem}
\usepackage{booktabs}

\newcommand{\dd}{{\rm d}}
\newcommand{\ee}{{\rm e}}
\newcommand{\ii}{{\rm i}}
\newcommand{\nablab}{{\boldsymbol{\nabla}}}

\newcommand{\paperI}{\citetalias{2020ApJ...903...40D}}
\newcommand{\paperII}{\citetalias{2021ApJ...914..118D}}

\DeclareSIUnit{\gauss}{G}
\DeclareSIUnit{\year}{yr}
\DeclareSIUnit{\erg}{erg}
\DeclareSIUnit{\eV}{eV}
\DeclareSIUnit{\parsec}{pc}
\newcommand{\gcc}{\gram\per\cubic\centi\meter}

\shorttitle{3D Models of Magnetar Outbursts}
\shortauthors{De~Grandis et al.}
\graphicspath{{./}{figures/}}

\begin{document}

\title{Three-Dimensional Magneto-Thermal Simulations of Magnetar Outbursts}

\correspondingauthor{Davide De Grandis}
\email{davide.degrandis@phd.unipd.it}

\author[0000-0001-5438-0908]{Davide De Grandis}
\affiliation{Department of Physics and Astronomy, University of Padova, via Marzolo 8, I-35131 Padova, Italy}
\affiliation{Mullard Space Science Laboratory, University College London, Holmbury St.\ Mary, Surrey, RH5~6NT, United Kingdom}

\author[0000-0003-3977-8760]{Roberto Turolla}
\affiliation{Department of Physics and Astronomy, University of Padova, via Marzolo 8, I-35131 Padova, Italy}
\affiliation{Mullard Space Science Laboratory, University College London, Holmbury St.\ Mary, Surrey, RH5~6NT, United Kingdom}

\author[0000-0002-1768-618X]{Roberto Taverna}
\affiliation{Department of Physics and Astronomy, University of Padova, via Marzolo 8, I-35131 Padova, Italy}

\author{Elisa Lucchetta}
\affiliation{Department of Physics and Astronomy, University of Padova, via Marzolo 8, I-35131 Padova, Italy}

\author[0000-0003-1044-170X]{Toby S.\ Wood}
\affiliation{School of Mathematics and Statistics, Newcastle University, Newcastle upon Tyne, NE1~7RU, United Kingdom}

\author[0000-0001-5326-880X]{Silvia Zane}
\affiliation{Mullard Space Science Laboratory, University College London, Holmbury St.\ Mary, Surrey, RH5~6NT, United Kingdom}



\begin{abstract}
The defining trait of magnetars, the most strongly magnetized neutron stars (NSs), is their transient activity in the X/$\gamma$-bands. In particular, many of them undergo phases of enhanced emission, the so-called outbursts, during which the luminosity rises by a factor $\sim 10$--$1000$ in a few hours to then decay over months/years. Outbursts often exhibit a thermal spectrum, associated with the appearance of hotter regions on the surface of the star, which subsequently change in shape and cool down. Here we simulate the unfolding
of a sudden, localized heat injection in the external crust of a NS with a 3D magneto-thermal evolution code, finding that this can reproduce the main features of magnetar outbursts. A full 3D treatment allows us to study for the first time the inherently asymmetric hot-spots which appear on the surface of the star as the result of the injection and to follow the evolution of their temperature and shape. We investigate the effects produced by different physical conditions in the heated region, highlighting in particular how the geometry of the magnetic field plays a key role in determining the properties of the event.

\end{abstract}


\defcitealias{2020ApJ...903...40D}{Paper I}\defcitealias{2021ApJ...914..118D}{Paper II}

\section{Introduction} \label{sec:intro}

With their huge magnetic fields ($B\sim\num{e13}$--$\SI{e15}{\gauss}$), magnetars are the most strongly magnetized objects in the present universe. Observationally identified with the Soft $\gamma$-repeaters and the Anomalous X-ray Pulsars (SGRs and AXPs, \citealp[e.g.][]{2001nsbh.conf..369T}), they are set apart from the other classes of isolated neutron stars (NSs) by their violent, transient high-energy activity, 
thought to be associated with the fast dissipation of large amounts of magnetic energy. Transient phenomena in magnetar sources fall into two broad categories: bursts/flares of relatively short duration (from $\lesssim\SI{1}{\second}$ to a few hours) and very diverse luminosities, reaching up to $\approx\SI{e47}{\erg\per\second}$ in the rare hyper-energetic giant flares, and much longer outbursts
\citep[see e.g.][for reviews]{2015RPPh...78k6901T,2017ARA&A..55..261K,2021ASSL..461...97E}. The latter are characterized by a sudden ($\approx\,$hours) increase of the flux (by a factor $\sim 10$--$1000$ with respect to the quiescent level), often accompanied by substantial modifications of the X-ray spectrum and of the pulse profile. These are related to an enhanced thermal emission, linked to the appearance of hotter region(s) on the surface of the star ($k_BT\sim 0.5$--$\SI{1}{\kilo\eV}$) and in some cases of a power-law tail in the $\sim 5$--$\SI{10}{\kilo\eV}$ range. The hot spot then shrinks and cools as the outburst decays on a timescale of $\sim\;$months/years \citep[see][for a comprehensive overview of outburst properties and the Magnetar Outburst Online Catalog\footnote{\url{http://magnetars.ice.csic.es}\label{note1}}]{2018MNRAS.474..961C}.

The association of outbursts with modifications in the surface thermal map has been taken as suggestive of some form of (magnetic) energy injection inside the NS crust \citep[e.g.][]{2011ApJ...727L..51P}, although the exact mechanism responsible for the heating is still debated (see Section \ref{sec:outbust_models}). This makes the study of magnetar outbursts an ideal ground for evolutionary magnetothermal simulations. The idea of using codes originally developed for treating NS cooling is not a new one and was exploited both in one spatial dimension, considering only the thermal evolution \citep[see e.g.][and references therein]{2021MNRAS.500.4491Y}, and in 2D for the coupled magnetothermal problem\defcitealias{2012ApJ...750L...6P}{PR+12}\citep[][in the following \citetalias{2012ApJ...750L...6P}]{2012ApJ...750L...6P}. 
In this work, we revisit the unfolding of an outburst as the result of a sudden heat injection in the magnetar crust. We take advantage of a full 3D treatment 
to elucidate the role played by the properties of the heated region and of the 
magnetic field in situations in which no spatial symmetries are present. In particular, this allows us to investigate how the position of the heated region with respect to the star magnetic field influences the outburst properties.

The paper is organized as follows. In Section \ref{sec:input} we review the input physics of our model and the main features of the numerical scheme used to compute the magnetothermal evolution. A set of simulations exploring the dependence of the outburst properties on several quantities characterizing the heat source and its location is presented in Section \ref{sec:outbust_models}. Finally, discussion follows in Section \ref{sec:discussion}.

\section{Input physics and numerical methods}
\label{sec:input}
In this work, the magneto-thermal evolution of a NS is computed following the same approach described in \citet{2020ApJ...903...40D} and \citet[][hereafter \paperI\ and \paperII, respectively]{2021ApJ...914..118D}, to which we refer for a more in-depth account. Here, after briefly reviewing the main features of our model, we focus on the major modifications introduced with respect to \paperI\ and \paperII, to deal most effectively with short-term phenomena. 

We study the magneto-thermal evolution of NSs in the crust only, under the assumption that (most of) the magnetic flux has been expelled from the superconducting core via the Meissner effect. The core of the star, then, is treated as a boundary condition in prescribing the temperature at the crust bottom.
The evolution of the crust is computed by solving the Hall induction equation in the electron MHD regime (eMHD, i.e. in the assumption that only electrons can move), and the heat balance equation
\begin{align}
 \frac{\partial\boldsymbol{B}}{\partial t} &= -c\,\nablab\times\left[\sigma^{-1}\boldsymbol{J} +\frac{1}{\ee cn_e}\boldsymbol{J}\times{\boldsymbol{B}}\right]
 \label{eq:Induction}
 \\
  C_v\frac{\partial T}{\partial t} &= - \nablab\cdot\left( 
  - \boldsymbol{k}\cdot\nablab T 
  \right) + \frac{1}{\sigma}\boldsymbol{J}^2-N_\nu \label{eq:heat}
\end{align}
where $\boldsymbol{B}$ is the magnetic field, $T$ is the temperature, $n_e$ is the electron density, $\mu= c\hslash (3\pi^2n_e)^{1/3}$ is the electron chemical potential, $\mathbf{J}=c\nablab\times\mathbf{B}/4\pi$ is the current density, $\boldsymbol{k}$ is the thermal conductivity tensor and $C_v$ is the heat capacity (per unit volume). The electrical and thermal conductivity $\sigma$ and $\boldsymbol{k}$ are taken as those of a strongly degenerate gas of electrons,
\begin{align}
\sigma&=\ee^2c^2\,\frac{n_e\tau(\mu)}{\mu}\,, 
\\
(\boldsymbol{k}^{-1})_{ij}&=\frac{3\ee^2}{\pi^2k_b^2T} 
\left(\frac{1}{\sigma}\delta_{ij}+\frac{\epsilon_{ijk}B_k}{\ee c n_e}\right)\,,
\label{eq:sigmakappa}\end{align}
where $\tau$ is the relaxation time and $\epsilon_{ijk}$ is the Levi-Civita symbol; $\hslash$ and $k_B$ are the reduced Planck and Boltzmann constants, and e is the electron charge. $N_\nu$ accounts for neutrino emission from weak processes in the crust; it depends strongly on $T$ and becomes active above $\sim\SI{3e9}{\kelvin}$ \citep[see \paperI\ and also][]{2001PhR...354....1Y}.
At variance with \paperI\ and \paperII, we decided to suppress the thermopower term in Equations (\ref{eq:Induction}) and (\ref{eq:heat}) since it is small in most situations and rather troublesome to treat numerically (see Section \ref{sec:discussion}).

As already mentioned, the internal boundary conditions (at the crust-core interface) are given by requiring that the field can not enter the core and specifying the core temperature evolution (which, nevertheless, occurs on much longer timescales than those considered here). The external one (at the crust-magnetosphere interface) for the magnetic field is that it should match a potential solution, $\nablab\times\boldsymbol{B}=0$, due to the negligible currents in the magnetosphere with respect to the crustal ones; the corresponding temperature boundary condition is given by specifying the thermal gradient according to the properties of the blanketing envelope. This is the geometrically thin but optically thick external layer that hosts a large temperature gradient; it is treated separately in the plane-parallel approximation and is included in our simulations as a relation between the temperature at the envelope bottom and at the surface, $T_b$ and $T_s$ in the form \citep{1983ApJ...272..286G, envelopes} \begin{equation}T_s(T_b, g, B, \Theta_B)=T_s^{(0)}(T_b, g)\,\mathcal{X}(T_b, B,\Theta_B)\label{eq:tsuruta}\end{equation} where $g$ is the surface gravity and $\Theta_B$ is the angle that the field forms with the normal to the surface; if not stated otherwise, we employed the expressions for $T_s^{(0)}$ as calculated in \citet{1983ApJ...272..286G} with the magnetic correction $\mathcal{X}(T_b,\mathbf{B})$ given in \citet{envelopes} for an iron envelope \citep[see also the review by][]{2021PhR...919....1B}. We note that these expressions were derived for use in standard cooling codes, where a thick envelope is prescribed (with a typical density at the base $\rho_b\approx 10^{10}\ \mathrm{g/cm}^3$), while in the present case a much thinner envelope is required. The properties of the envelope depend on $\rho_b$ \citep[see e.g.][]{2016MNRAS.459.1569B} but unfortunately no calculations are available in the literature, to our best knowledge, for $\rho_b\approx 10^{7}\ \mathrm{g/cm}^3$ and in the magnetized case. Ideally one should compute an envelope structure tailored on
the problem at hand. However, given the main focus of the present investigation, we stick to our previous choices, after having checked that our results are not much sensitive to the
details of the envelope, as we show in Section \ref{sec:discussion}.
General-relativistic corrections are neglected, since we are treating a thin layer; they are nevertheless expected to have at most only a quantitative rather than qualitative impact on the results \citep{2009A&A...496..207P}. For this reason, all the quantities we derive are expressed in the star local rest frame but they can be easily transformed into those measured by an observer at infinity (e.g.\ for the luminosity it is $L_\infty = (1-2GM/R c^2)\,L$, assuming that the space-time around the NS is described by the Schwarzschild metric, with $M$ and $R$ the mass and radius of the star, respectively).

This system of equations, written in dimensionless form, was then numerically solved in three spatial dimensions by means of the {\sc parody} code \citep{dormy} which employs a pseudo-spectral scheme based on spherical harmonics; {\sc parody} was first adapted to the study of NS magnetic evolution alone \citep{2015PhRvL.114s1101W} and then generalized to deal with the coupled magneto-thermal problem in \paperI\ \citep[see also][]{2020NatAs.tmp..215I}. 

In this work, we employ an updated version of the code that has been specifically developed for a more realistic treatment of short-term phenomena, like those which occur following a sudden heat deposition in the outermost layers of a NS. 
In dealing with the secular magneto-thermal evolution, in fact, the rather long runs, extending up to an Ohm time ($\approx 10^6$ yr), forced us to introduce some assumptions to avoid exceedingly large computational times. To this end, we had specified a rather high cutoff density in the crust ($\approx\SI{e10}{\gram\per\cubic\centi\meter}$), implying the presence of a quite thick blanketing envelope, not to be limited by too small timesteps. Furthermore, we used a simplified expression for the specific heat that allowed us to treat the heat flux propagation with an implicit time-advance scheme; specifically, $C_v$ was taken to be linearly dependent on the temperature, which is a good description for the (subdominant) electron contribution only, to ensure the efficiency of the numerical scheme. This is a reasonable choice in treating the long-term evolution, since the term $\propto\partial T/\partial t$ is strongly suppressed, but it becomes questionable when impulsive phenomena are considered.

These two approximations were then relaxed. First, the integration domain was extended to include the outer crustal layers, which were treated as part of the blanketing envelope in our previous studies. This was done by using the model proposed by \citet{2021MNRAS.500.4491Y}, who provide a simple scaling for the dimensionless Fermi momentum $x_r=p_F/mc\propto n_e^{1/3}$ as a function of the depth $z$ as
\begin{equation}
    x_r^3=\left[\frac{z}{z_0}\left(2+\frac{z}{z_0}\right)\right]^{3/2}\,;\quad z_0=\frac{Zm_ec^2}{m_ug_sA}\approx\SI{10}{\meter}\label{eq:EoSYak20}
\end{equation}
where the scale height $z_0$ is given in terms of the atomic mass unit $m_u$ and of the (local) gravitational acceleration $g_s$. In the following, standard NS parameters will be used ($M=1.4\,M_\odot$, $R=\SI{12}{\kilo\meter}$, $g_s=\SI{1.59e14}{\centi\meter\per\squared\second}$), as well as a constant Fe composition ($Z=26$, $A=56$). For consistency, the density profile of the inner crust used in \paperI\ and \paperII \ \citep[taken from][]{2014MNRAS.438.1618G} has been retained, with the outer crust built above it for densities $\rho \lesssim \SI{e11}{\gram\per\cubic\centi\meter}$. The transition point was chosen with the additional requirement that the two profiles join smoothly. Note, however, that the inner crust plays no role in the outburst physics, since it virtually does not evolve on the relevant timescales and the details of its EoS, as well as the exact depth of the matching point between the two profiles, are hence irrelevant \citep[as already noted by][themselves]{2021MNRAS.500.4491Y}. This additional layer extends down to $\rho\simeq\SI{e6}{\gcc}$, implying a rather thin blanketing envelope. As in \paperI\ and \paperII, the electron relaxation time $\tau$ has been set to the constant characteristic value $\SI{9.9e-19}{\second}$, so that $\sigma\sim n_e^{-2/3}$; a more realistic treatment of this quantity is deferred to a future work (see Section\ \ref{sec:discussion}).

Second, we choose to adopt a minimal model for the specific heat which, nevertheless, accounts for both the lattice and the electron contributions \citep[e.g.][]{transport}
\begin{equation}
    C_v=C_v^\text{lattice}+C_v^\text{el}=3k_Bn_i+\frac{\pi^2k_B^2}{m_ec^2}\frac{\sqrt{x_r^2+1}}{x_r^2}T
\end{equation}
where $n_i=Zn_e$ because of charge balance. The first, temperature-independent term describes a classical bcc lattice and is the dominant one up to very high temperatures (where the classical expression also gets inaccurate). The second term describes the contribution of strongly degenerate electrons. Since the two terms have different temperature dependencies, no manipulation, such as writing the equation in terms of $T^2$ as done before, can eliminate the $T$ dependence in the LHS of Equation (\ref{eq:heat}). Using again an implicit integration scheme with the new expression for $C_v$ would require to invert the time advance operator at every time step making the computational time unacceptably long. Instead, the whole temperature equation has been moved to the explicit part of the time advance algorithm, and treated with the Adams-Bashforth scheme. This is still more costly than using the implicit algorithm, but it does not significantly increase the computational time per time step. The magnetic evolution scheme remained unchanged.

Having adopted an explicit scheme for the temperature equation makes, however, the numerical handling of neutrino losses challenging. Neutrino emission strongly depends on temperature ($N_\nu\propto T^7$--$T^8$) and computing it consistently up to very high $T$ may require extremely small time steps. In this case, the evaluation of some terms may be strongly limited by machine accuracy, with the result that they can not get properly updated. To avoid this, the time stepping algorithm has been left unchanged and a numerical check imposed on the neutrino loss term instead. If the time step is too large the neutrino emission can be overestimated and \emph{overshoot}, resulting in a negative temperature in correspondence of what was a very hot point. This has been prevented by requiring that at each time iteration the temperature of any given point could not decrease by more than $50\%$ of its original value. This mimics the fast cooling by neutrinos but spreading it over a more manageable time, which is nevertheless limited to a few time steps, as we checked a posteriori.

\section{Outburst Models}
\label{sec:outbust_models}

There is currently no clear picture as to what the exact mechanism causing the heating of the observed hot spot(s) may be. It has been suggested that the surface of the star is heated by returning currents or by some other dynamics of the field in the lower magnetosphere \citep[e.g.][]{2009ApJ...703.1044B,10.1093/mnras/stz1507}. Alternatively, some kind of fast magnetic energy dissipation may take place in the outermost layers of the crust, triggered by e.g.\ a mechanical failure due to magnetic stresses deeper inside \citep[e.g.][]{2011ApJ...727L..51P,2016ApJ...824L..21L,2019MNRAS.486.4130L,2021MNRAS.506.3578G}. 

In this section, a set of simulations intended to model magnetar outbursts will be presented. To keep clear of all the uncertainties concerning the exact mechanism triggering the outburst, we just assume that a given amount of energy is deposited in a localized patch of the outer crust, following the same approach outlined in \citetalias{2012ApJ...750L...6P}. In order to match the observed rise times ($\lesssim\SI{1}{\day}$), the heating phase should be quite short, lasting much less than the overall duration of the event. As already noted by \citetalias{2012ApJ...750L...6P}, the actual duration of the injection phase, $\Delta t$, does not affect the subsequent evolution as long as the total (time integrated) heat $H$ is the same and $\Delta t$ is not much longer than the characteristic heat diffusion time across the crust, in which case a quasi-stationary state develops. For the cases considered here, it has been checked that $\Delta t\lesssim\SI{1}{\hour}$ yields indistinguishable results. The injection was done over several timesteps and the introduction of a modulation profile in time has been verified to produce no effect as well.

Heat was injected into a spatially localized region in the external crust ($\rho\approx\SI{e7}{\gram\per\cubic\centi\meter}$) with a thickness of $\approx\SI{100}{\meter}$ and a diameter of some kilometers to reproduce the observed hot spot areas. Consequently, we added to Equation (\ref{eq:heat}) a heating term in the form
\begin{equation}
    H'=\dot H s(r,\theta,\phi)
    \label{eq:heat_injec}
\end{equation}
where $\dot H$ is the heat injection rate per unit volume and $s(r,\theta,\phi)$ 
is a smoothing function introduced to avoid sharp gradients. Its actual form is
\[\sin^4\left(\!\pi\,\frac{r_2-r\hphantom{_1}}{r_2-r_1}\right)\cdot\sin^4\left(\!\pi\,\frac{\theta_2-\theta\hphantom{_1}}{\theta_2-\theta_1}\right)\cdot\sin^4\left(\!\pi\,\frac{\phi_2-\phi\hphantom{_1}}{\phi_2-\phi_1}\right)\]
in a region specified by the three intervals $[r_1,r_2]$, $[\theta_1,\theta_2]$ and $[\phi_1,\phi_2]$. We took a shallow layer $\Delta r\approx\SI{100}{\meter}$ extending some kilometers across.
Then, the total amount of heat added is then
\begin{equation}
    H=\Delta t \int_V\dot H\, s(r,\theta,\phi)\,r^2\sin\theta\,\dd r \,\dd\theta\, \dd\phi
    \label{eq:heat_tot}
\end{equation}
and the integral is evaluated over the entire heated volume.
Note, however, that since crustal neutrino emission is very non-linear and much stronger in the hotter zone at the center of the patch, the temperature profile tends to be flattened out, so that the exact form of the smoothing profile is not crucial. 

\subsection{Neutrino saturation} 
\label{sec:saturation}
A major effect of crustal neutrino emission, already discussed by \citetalias{2012ApJ...750L...6P} in 2D simulations, is the saturation of the maximum (photon) luminosity for increasing values of the total heating input. 
The strong temperature dependence of neutrino losses implies, in fact, that as the total heat input $H$ grows, keeping all the other parameters fixed, a greater and greater fraction gets quickly released in the form of neutrinos, without contributing to the thermal emission.

In order to assess if and how this picture changes within a 3D approach we performed a number of runs taking as a background a NS with an initial crustal temperature of $\SI{e8}{\kelvin}$ and a dipolar field of $\SI{e13}{\gauss}$. The field has again been built as the poloidal part of a force free field. It has also been checked that the inclusion of a toroidal component, even much stronger than the poloidal one, has virtually no direct effect on the outburst, since it is confined in the deeper layers by the requirement of negligible magnetospheric currents (the boundary condition $\nablab\times\boldsymbol{B}\vert_{R_\star}=0$). The heating term was activated few time steps after the beginning of the simulation, so that the magnetic configuration is virtually the same as the initial one.

The heated patch is located at $\rho\simeq\SI{3e7}{\gram\per\cubic\centi\meter}$ and covers a rather large area ($\sim20\%$ of the surface) centered at $\ang{26}$ north of the magnetic equator. Albeit not completely realistic, this has been purposely done for two reasons. The first, of more practical nature, is that in order to observe the saturation very large energy inputs are required, implying steep thermal gradients that are  difficult to handle within a spectral scheme; a (relatively) large  heated region results in smoother, numerically manageable gradients. Moreover, as it will be discussed in detail in Section \ref{sec:mag_geometry}, the location of the heated region with respect to the magnetic field geometry is important, so that the use of a large patch has been preferred in order to average out these effects.

\begin{figure}
    \centering
    \includegraphics[width=.49\textwidth]{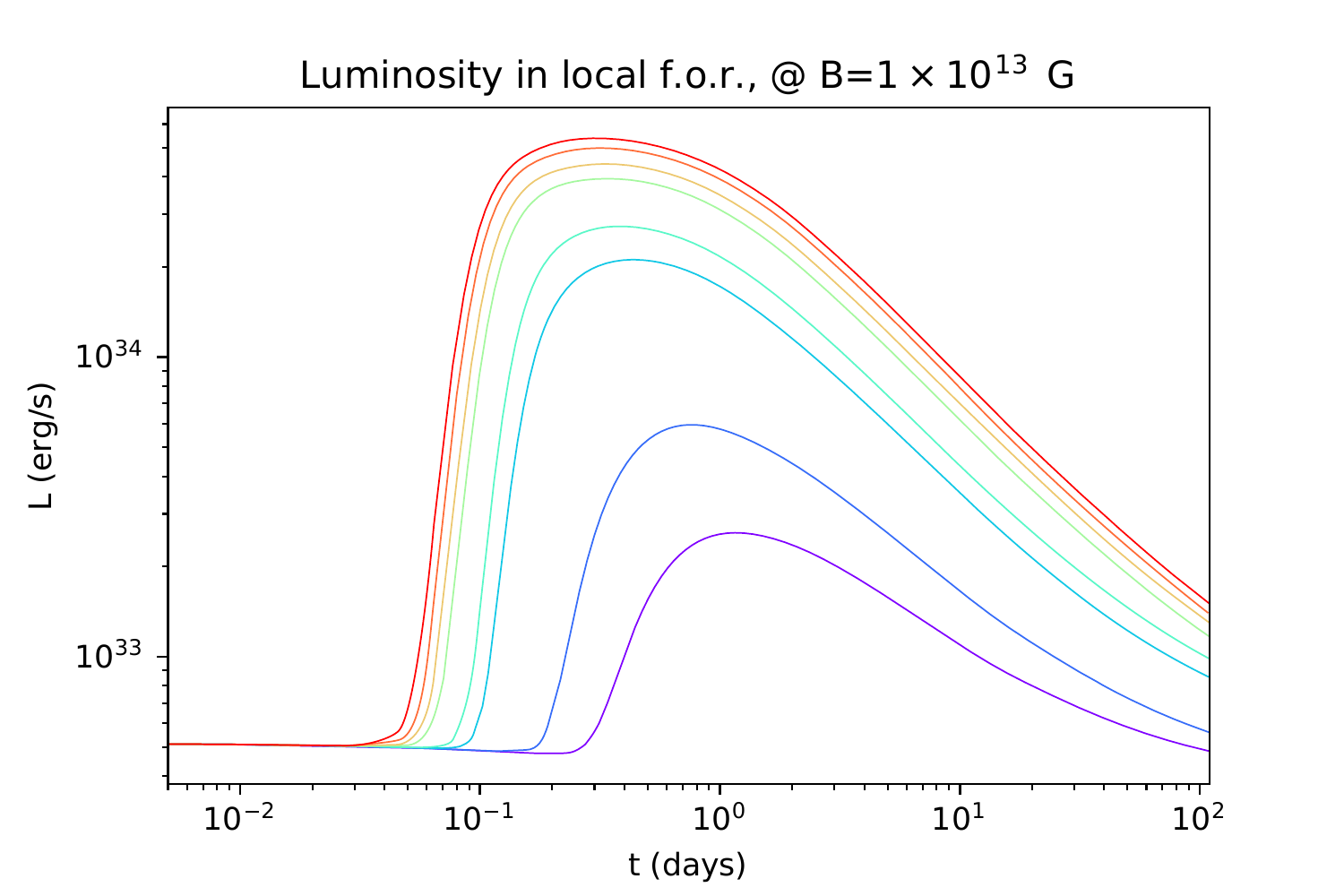}
    \caption{Evolution of luminosity after the injection of increasing amounts of heat; higher curves correspond to larger $H$ (see Figure\ \ref{fig:saturationProfiles} for the actual values) for a NS with $T(0)=\SI{e8}{\kelvin}$, $B\sim\SI{e13}{\gauss}$.}
    \label{fig:saturationCurves}
\end{figure}

\begin{figure}
\centering
\subfigure{\includegraphics[width=.45\textwidth]{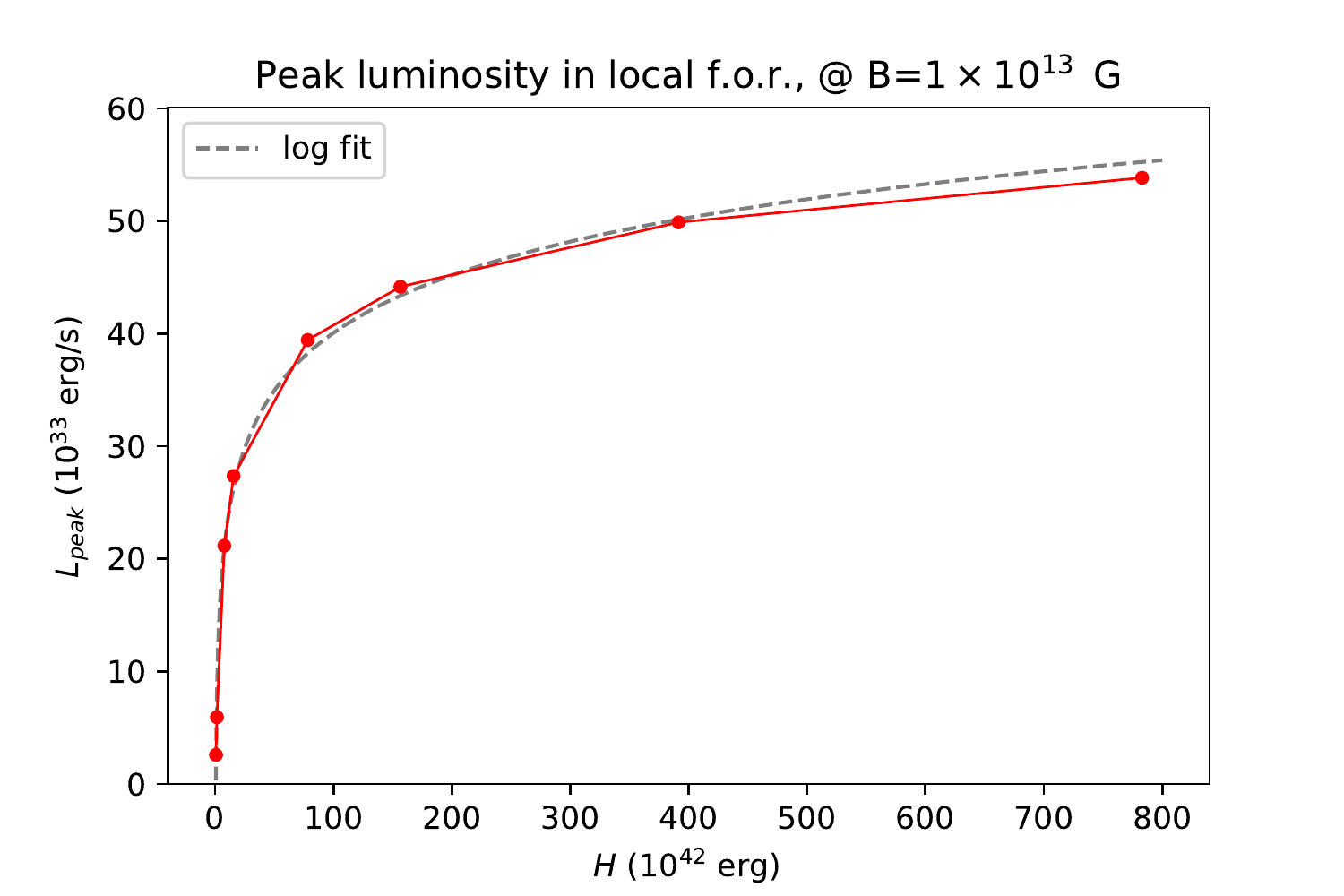}}
\subfigure{\includegraphics[width=.45\textwidth]{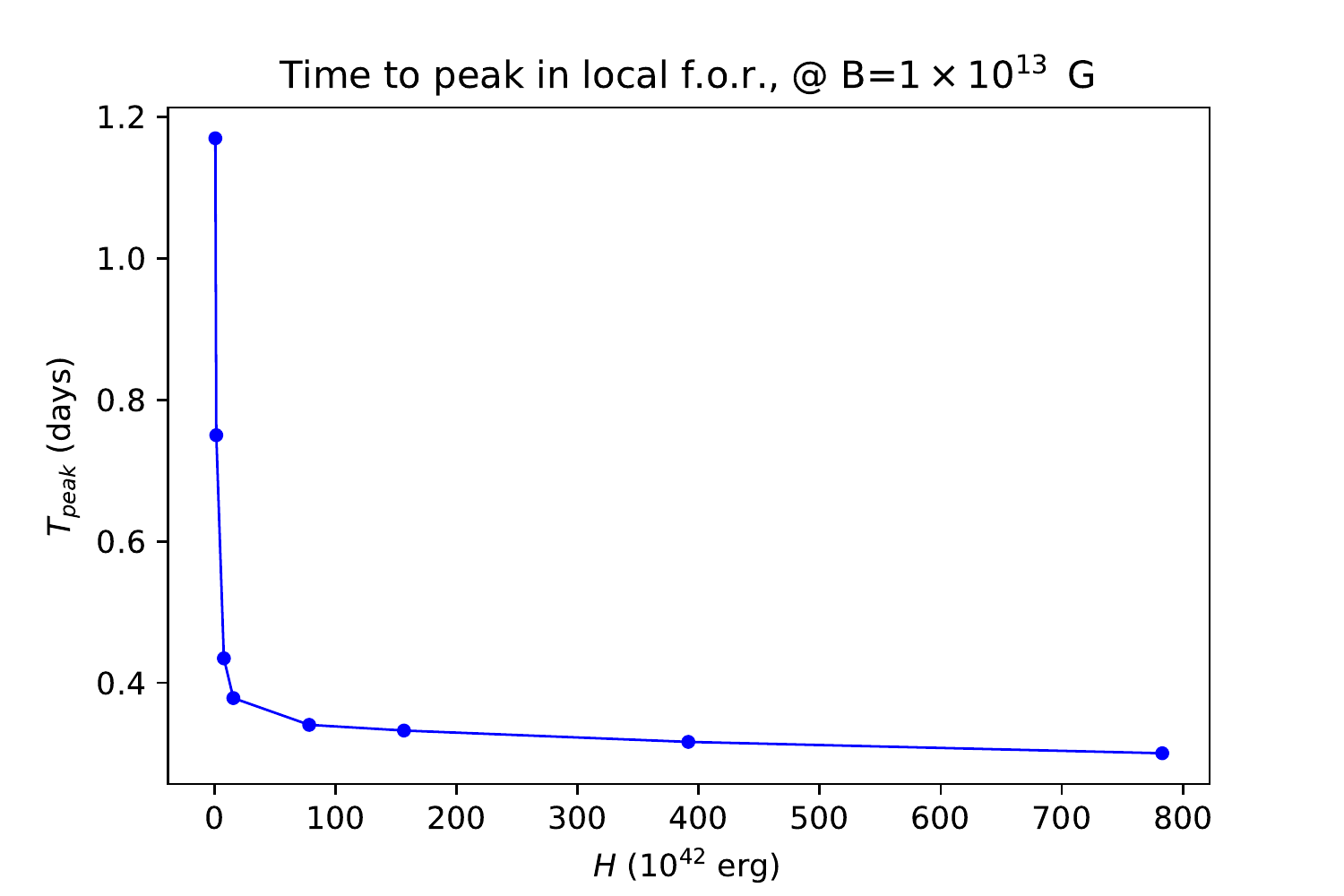}}
\caption{Peak luminosity (upper panel) and rise-time (lower panel) as functions of the heat injection for the curves in Figure\ \ref{fig:saturationCurves}. In the upper panel, a logarithmic fit to the points has been added as a reference.} \label{fig:saturationProfiles}
\end{figure}

Figure\ \ref{fig:saturationCurves} shows the evolution of the luminosity (as measured by a stationary observer at the surface of the star, the local f.o.r.\,) for different values of the heat input in a range $\SI{7e39}{\erg}\lesssim H\lesssim\SI{4e42}{\erg}$; here and in the following the luminosity has been computed assuming patch-wise blackbody emission at the local surface temperature. The general behavior is qualitatively similar to what is observed in real sources and coherent with the curves in \citetalias{2012ApJ...750L...6P}: a rise time of hours to $\sim\SI{1}{day}$, faster as $H$ increases (and hence the peak gets higher) and an overall duration $\gtrsim\SI{100}{\day}$, longer for the brighter cases. This time span is too short for the field to undergo any significant evolution.
The general shape of the curves does not change much between the various cases and, upon visual inspection, the curves appear to pile up as $H$ increases. In order to get a more quantitative evidence of this effect, the peak luminosity and the time it takes to reach the peak are plotted against the total injected heat $H$ in Figure\ \ref{fig:saturationProfiles}. The saturation effect, then, becomes evident: when the heating gets larger than $\approx\SI{e44}{\erg}$, the luminosity increase becomes very small and the rise-time does not change anymore.


\subsection{Dependence on the injection density} 
\label{sec:injec_depth}
\begin{figure}
    \centering
    \includegraphics[width=.49\textwidth]{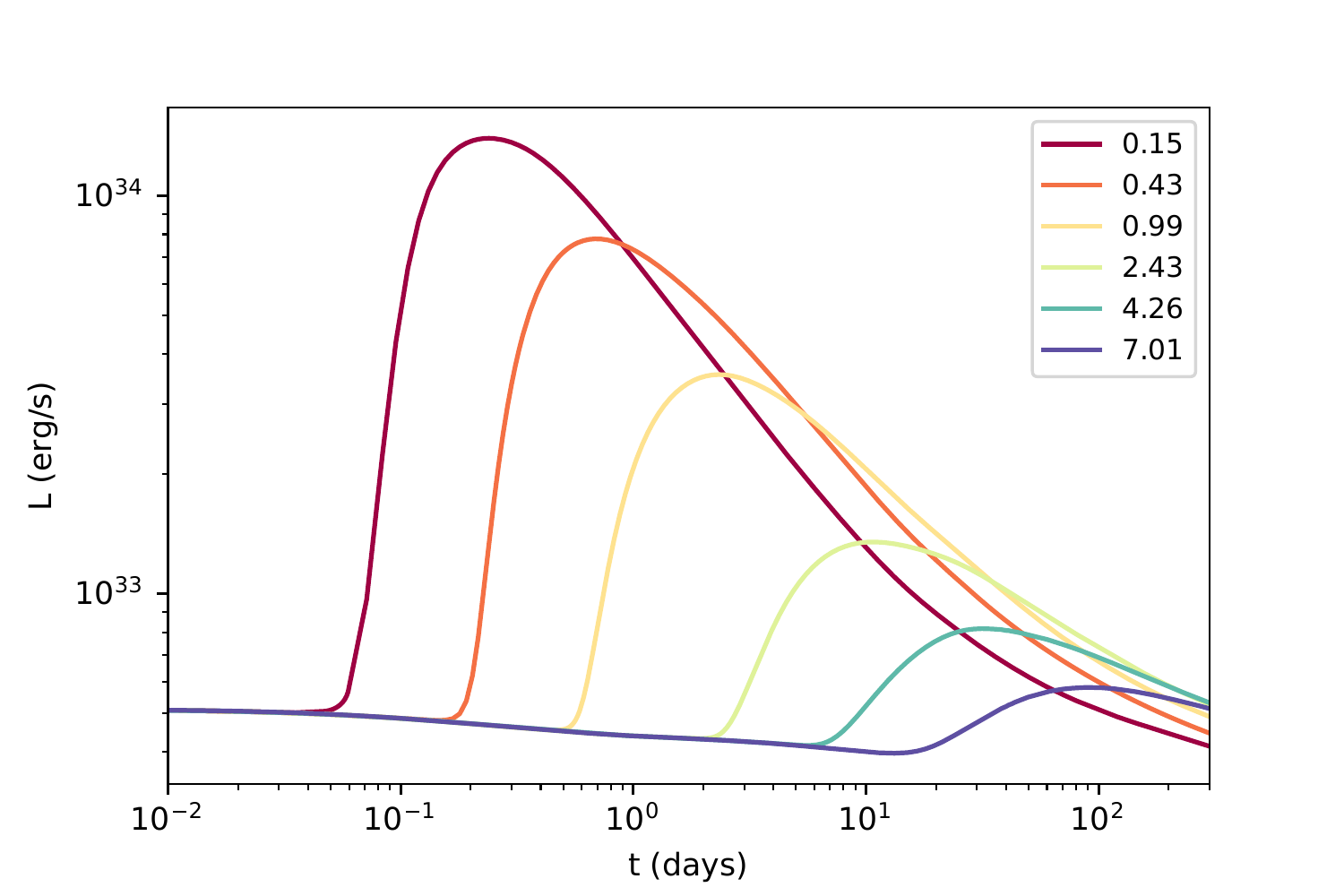}
    \caption{Lightcurves corresponding to the injection of the same amount of heat, $H\simeq\SI{2.9e40}{\erg}$, at different depths. The curves are color-coded according to the value of the average density in the injection region in units of $\SI{e7}{\gcc}$.}
    \label{fig:depth}
\end{figure}

\begin{figure}
\centering
\subfigure{\includegraphics[width=.45\textwidth]{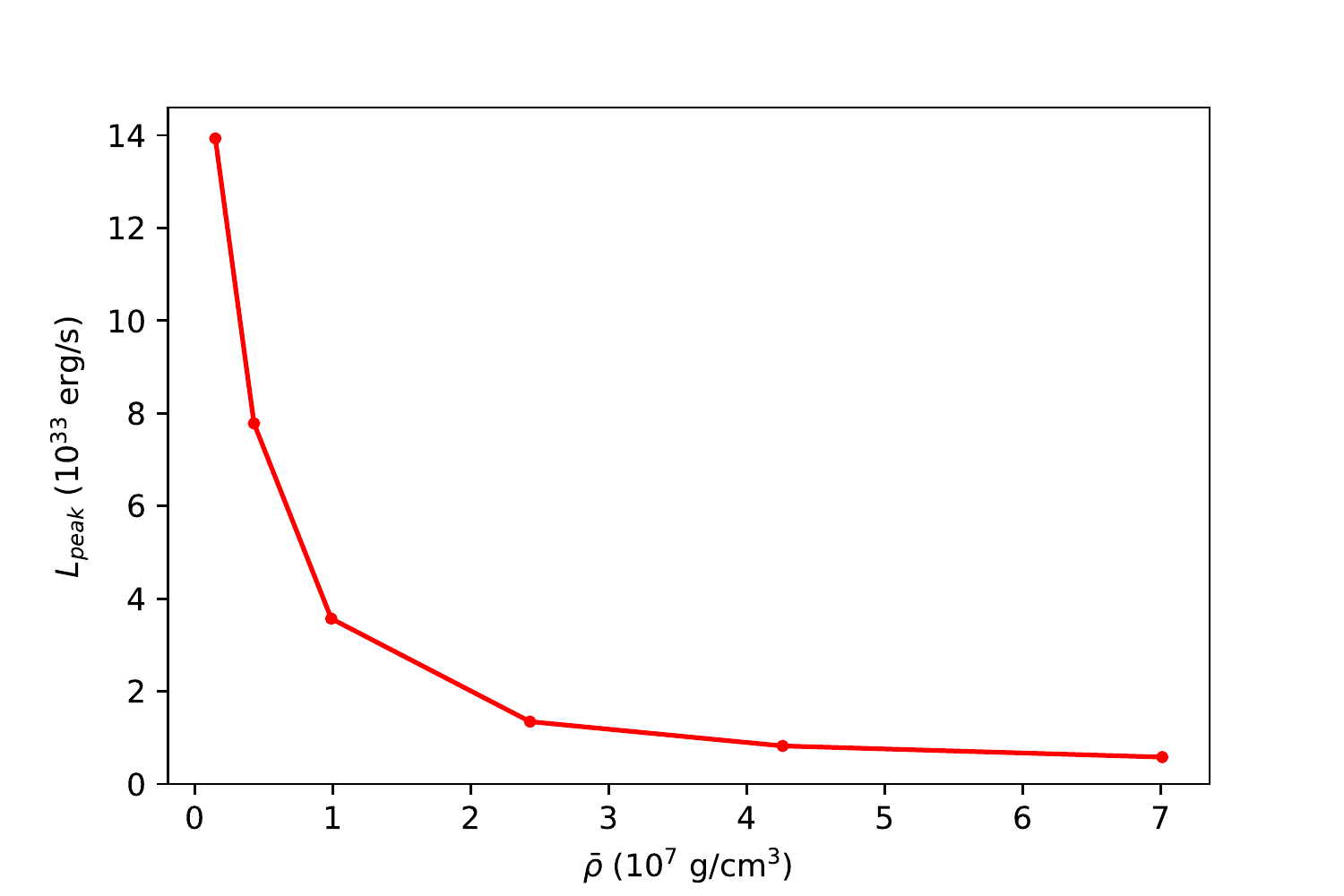}}
\subfigure{\includegraphics[width=.45\textwidth]{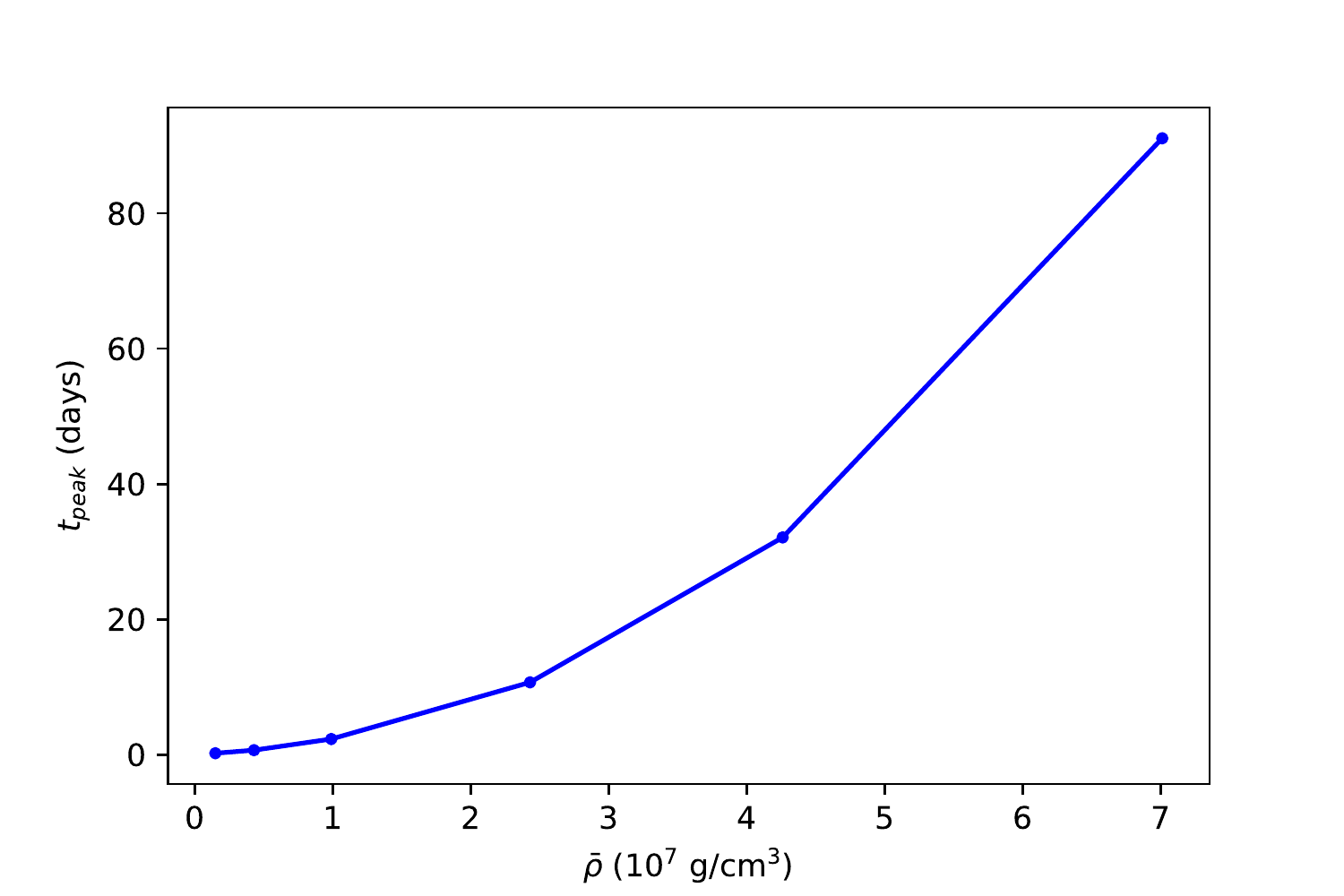}}
\caption{Peak luminosity (upper panel) and rise-time (lower panel) for the curves of Figure \ref{fig:depth} as a function of the average density of the injection region.
} \label{fig:depthProfiles}
\end{figure}

The cases considered so far assumed that heat was injected at the same depth in the crust. However, injecting the same amount of heat at different depths is expected to produce different results, since in deeper layers the density (and hence the specific heat) increases and heat needs to travel longer to reach the surface and modify the luminosity. Figure \ref{fig:depth} shows the lightcurves corresponding to the same injection of $H\simeq\SI{3e40}{\erg}$ in the same shallow patch of radius $\simeq\SI{2}{\kilo\meter}$ at different depths in the outer crust.
Since the density profile is quite steep ($\sim z^3$, see Equation \ref{eq:EoSYak20}), the shape of the lightcurves themselves turns out to be quite sensitive to the injection depth. In fact, as the depth increases the rise-time becomes longer and the peak lower, as expected (see Figure\ \ref{fig:depthProfiles}). Note that the luminosity decrease before the onset is not linked to the outburst iteself but rather to the (overall) cooling of the NS. This effect is in part degenerate with varying the total energy input $H$ at the same depth. Given that deep injection is radiatively inefficient and that the heat input can not be increased indefinitely, due to neutrino saturation, the outbursts likely originate from energy deposition in the outermost layers of the crust.
On the other hand, heating in the deep layers may indeed reach the surface, but too slowly and in too little an amounts to cause a proper outburst, and may produce just a moderate increase in luminosity, that would be confused with a variability of the quiescence flux.


\subsection{Dependence on the magnetic field strength} 
\label{sec:mag_strength}
\begin{figure}
    \centering
    \includegraphics[width=.51\textwidth]{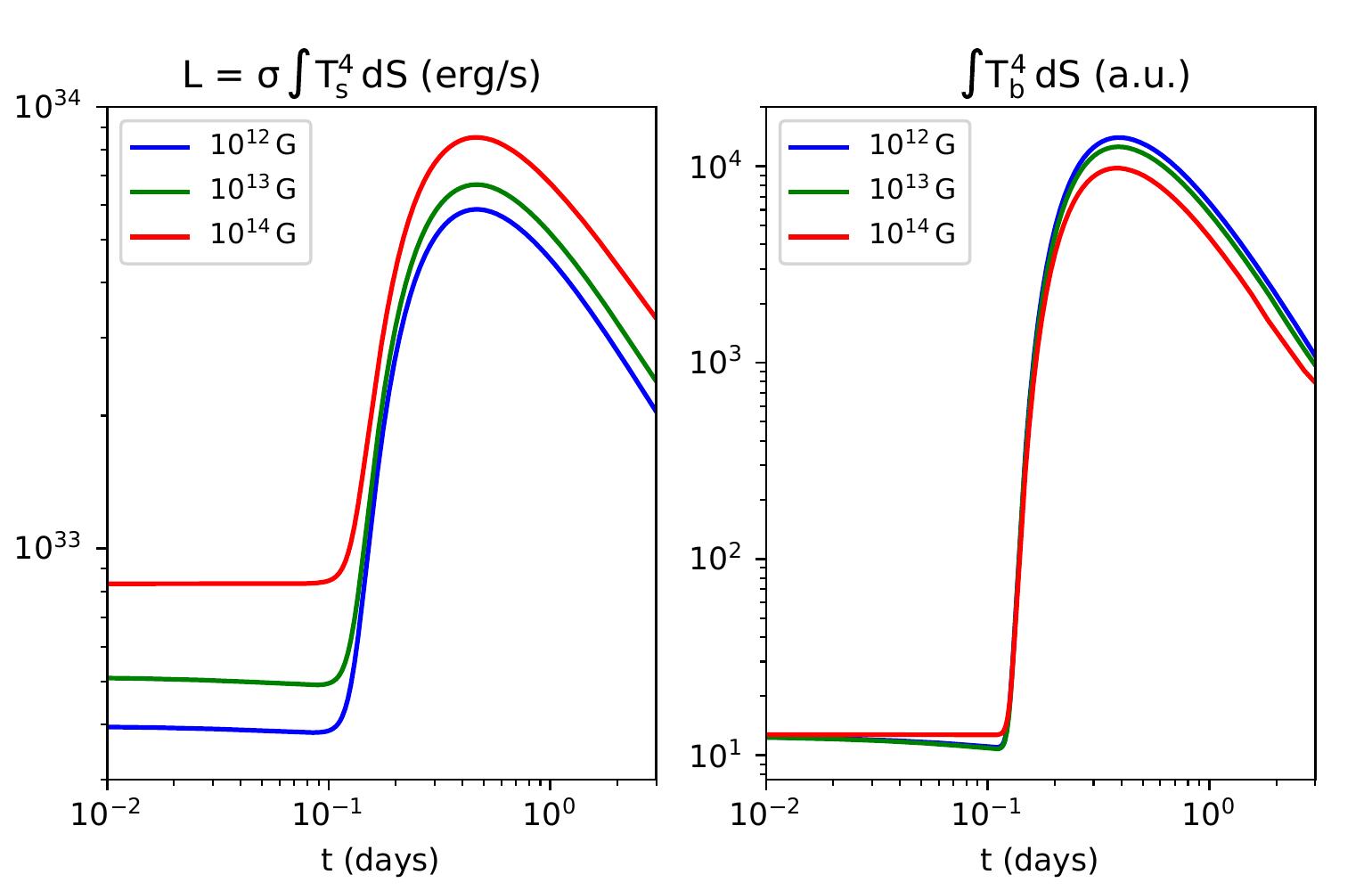}
    \caption{Evolution of the luminosity (left) and of the surface integral of the top-of-the-crust temperature to the fourth (right) in the first stages of the outburst for three different strengths of the (dipolar) magnetic field; all the other parameters are the same.}
    \label{fig:differentB}
\end{figure}

As a further step, we consider the dependence of the outburst evolution on the magnetic field. This actually involves both the strength and direction of the field lines, which (for a fixed field geometry) reflects into the position of the heated region. Let us consider a heat injection of $H\sim\SI{3e40}{\erg}$ in a $\sim\SI{5}{\kilo\meter}$ radius patch, in a region around the magnetic pole (slightly off-centered), again at a depth where $\rho\sim\SI{3e7}{\gcc}$ (we checked that considering either the north or south polar regions yield identical results). The resulting lightcurves are shown in Figure\ \ref{fig:differentB} (left panel) for a dipolar field of strength in the range \num{e12}--\SI{e14}{\gauss}. 
The different strength of the magnetic field influences both the outburst itself and the thermal evolution of the background NS. In particular, the equatorial region gets heated by the field to a varying degree, so that the luminosity of the more magnetized NSs has an additional contribution coming from the equatorial region. For this reason, in order to focus on the (polar) heated patch the evolution of the background has been subtracted out from the luminosity curves. In all these runs the time-step has been kept fixed at a small value ($\Delta t \sim\SI{1}{\minute}$) in order to avoid any potential inconsistencies caused by the anti-overshooting algorithm (see Section \ref{sec:input}), given that the adaptive time step built in the code depends on the field strength. The evolution of the luminosity, in particular the rise and peak times, show no significant variation and the overall shape is almost identical. As the magnetic field increases between the different models, the peak luminosity increases, as well. This fact may seem somewhat counter-intuitive, since the magnetic field acts as a thermal insulator (see Eq.\ \ref{eq:sigmakappa}), so that one might expect a lower peak luminosity for stronger fields. To better understand this effect, we compare the evolution of the luminosity to that of the quantity $\int T_b^4 \dd S$ (Fig.\ \ref{fig:differentB}, right panel). This just is a way to follow  the evolution of the temperature at the top of the crust, $T_b$, i.e. before calculating the actual surface temperature via Eq.\ \ref{eq:tsuruta}. Variations among the different curves are even smaller for this quantity and, as expected, the models with stronger fields yield lower peaks. Therefore, the hierarchy inversion is caused by the form of the $T_b$--$T_s$ relation we assumed. In particular, this reflects the fact that in the heat-blanketing envelope the longitudinal heat conductivity, which drives the heat transport near the pole, gets amplified by electron quantization effects; this makes the envelope polar regions more transparent to heat as the field grows \citep[the effect is the opposite near the equator, see][in particular their Figure 22]{2021PhR...919....1B}.

Since the shape of the lightcurves turns out to be not very sensitive to the field strength, in the following section we take as a representative value   $B\simeq\SI{e13}{\gauss}$; this choice is also more convenient on a computational ground because stronger fields exact a heavier computational toll.
\medskip

\subsection{Dependence on the magnetic field geometry} 
\label{sec:mag_geometry}
We now turn to the effects of the field geometry in the injection region. To this end, the same patch of the previous run was considered, this time with an injected energy of $H=\SI{4e40}{\erg}$, but at two different locations, either centered at the magnetic pole, or crossing the equator
. These two cases are representative of a geometry in which the field lines are mostly ``open'' (i.e. perpendicular to the surface) or ``closed'' (parallel to it), respectively. Furthermore, alongside the iron envelope considered so far, we explored a different composition, namely  the ``fully accreted'' envelope model from \cite{2021PhR...919....1B}. Our choice is in the spirit of bracketing the effects of different envelope chemical compositions by considering two limiting cases, although we are aware that a light element composition is not very likely in the case of an isolated magnetar. Moreover, light elements would likely be burnt by nuclear reactions during the outburst, providing an additional source of heat, as well as modifying the envelope properties as the event proceeds (which would also alter the new quiescence luminosity
); these effects are not considered in the present work. Nevertheless, envelopes with lighter compositions may be required to explain the huge luminosity variations of the most luminous outbursts. The accreted envelope is more transparent, so that the luminosity is higher for the same $H$, as expected (see also \citetalias{2012ApJ...750L...6P}). Moreover, in this case a somewhat longer time ($\lesssim\SI{1}{\year}$) is required to establish a thermal gradient across the crust, so that the injection was performed after this phase, which is of almost negligible duration (hours) with the more blanketing envelope.

\begin{figure}
    \centering
    \includegraphics[width=.49\textwidth]{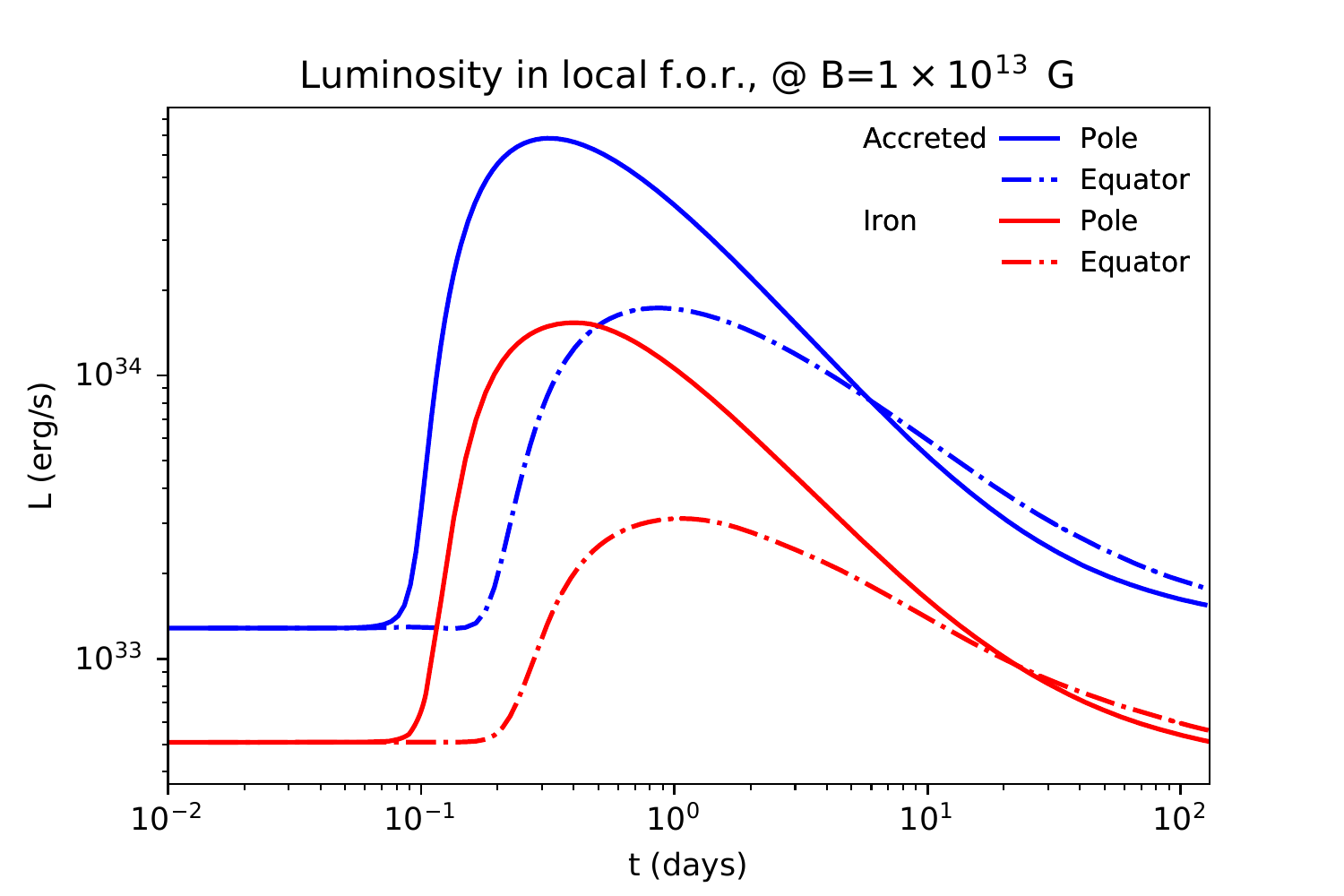}
    \caption{Outburst evolution for a NS with $B=\SI{e13}{\gauss}$ and an injection of $H\simeq\SI{4e40}{\erg}$ in two different positions with respect to the dipolar field, around the pole (continuous lines) and across the equator (dot-dashed lines). Two different envelope compositions are considered, a Iron one (red curves) and a light-element, accreted one (blue curves). The background evolution of the NS has been subtracted out.}
    \label{fig:envelope_bursts}
\end{figure}

\begin{figure}
\centering
\subfigure{\includegraphics[height=.098\textwidth]{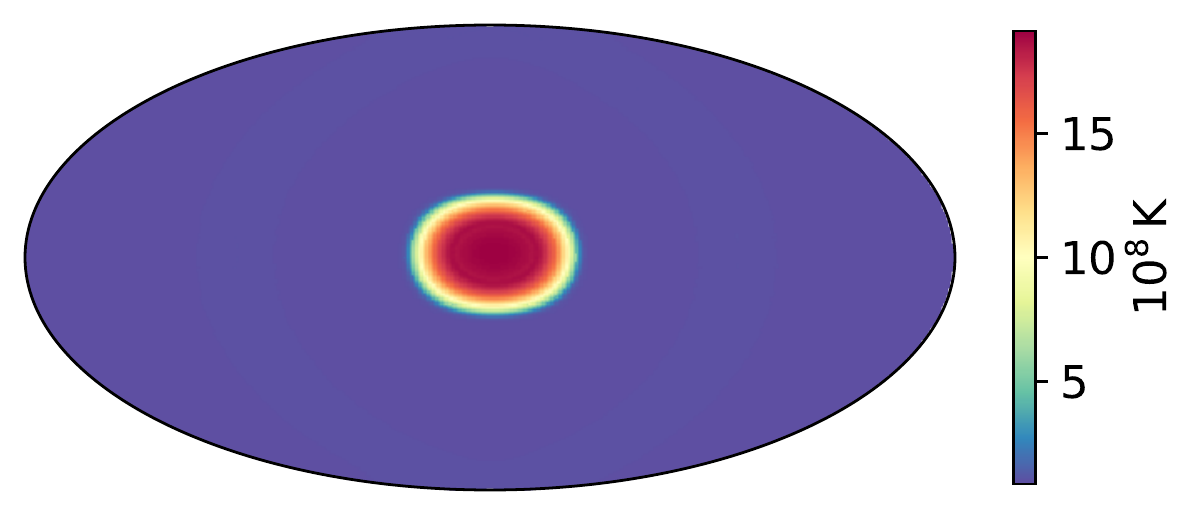}}
\subfigure{\includegraphics[height=.098\textwidth]{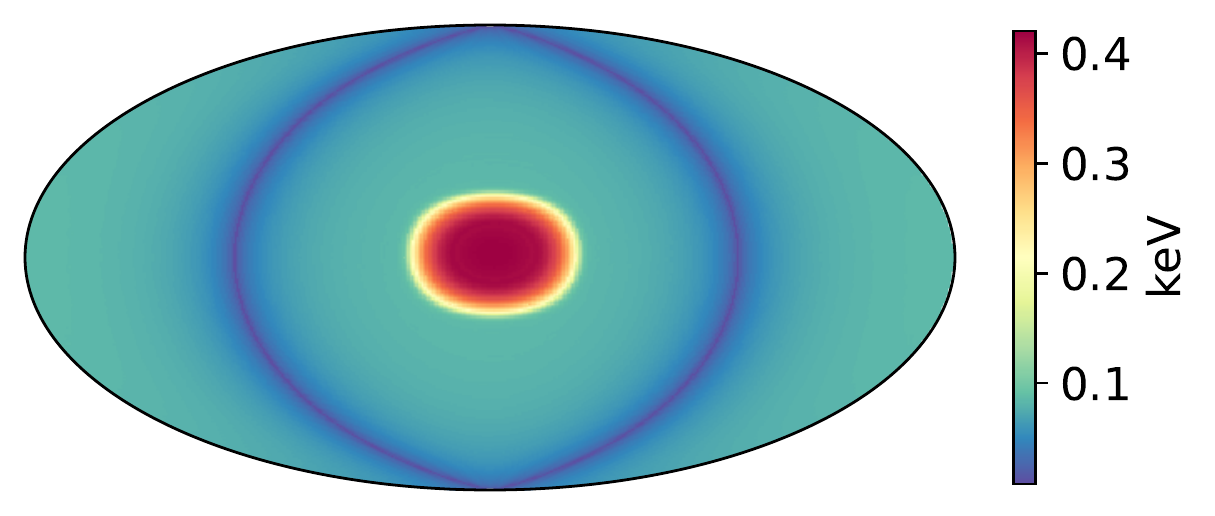}}\\
\subfigure{\includegraphics[height=.098\textwidth]{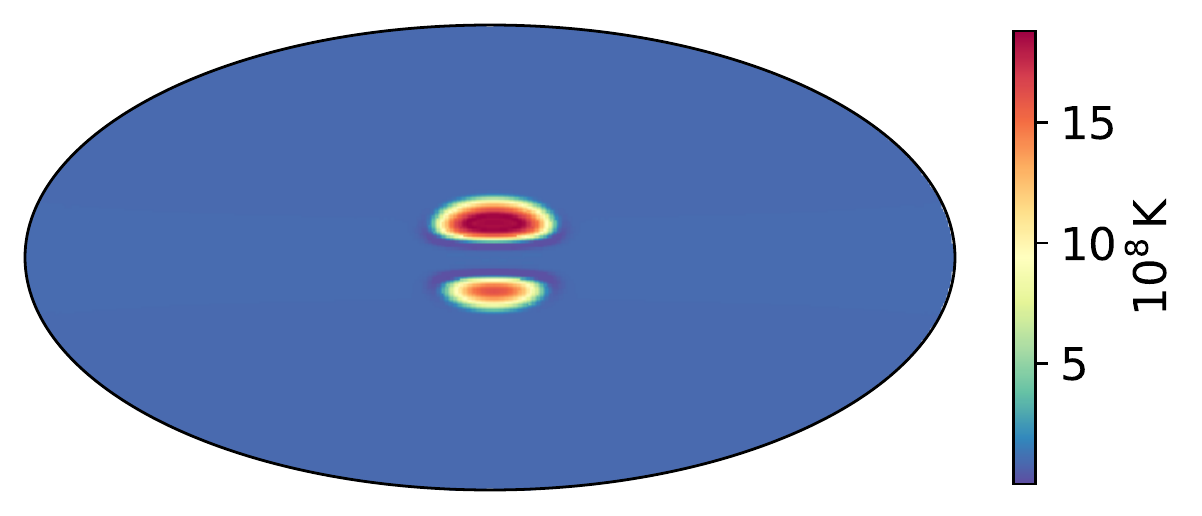}}
\subfigure{\includegraphics[height=.098\textwidth]{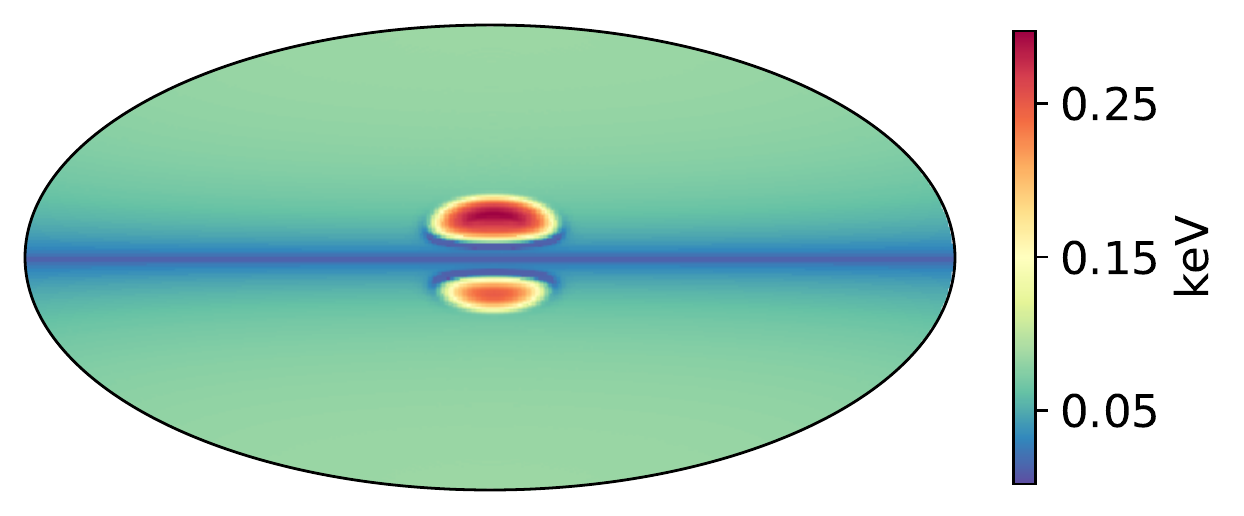}}
\caption{Thermal maps for the cases shown in Figure\ \ref{fig:envelope_bursts} after $t=\SI{9}{\hour}$ (rise phase) for the Iron envelope. The top row shows the case of polar injection (magnetic pole towards the observer), the bottom one refers to equatorial injection (magnetic pole upwards); the left-hand panels show the crustal temperature $T_b$, the right-hand ones the surface temperature $T_s$. The injection is slightly off-centered, hence the north-south asymmetry.} \label{fig:spot_shapes}
\end{figure}

\begin{figure}
\centering
\subfigure{\includegraphics[width=.45\textwidth]{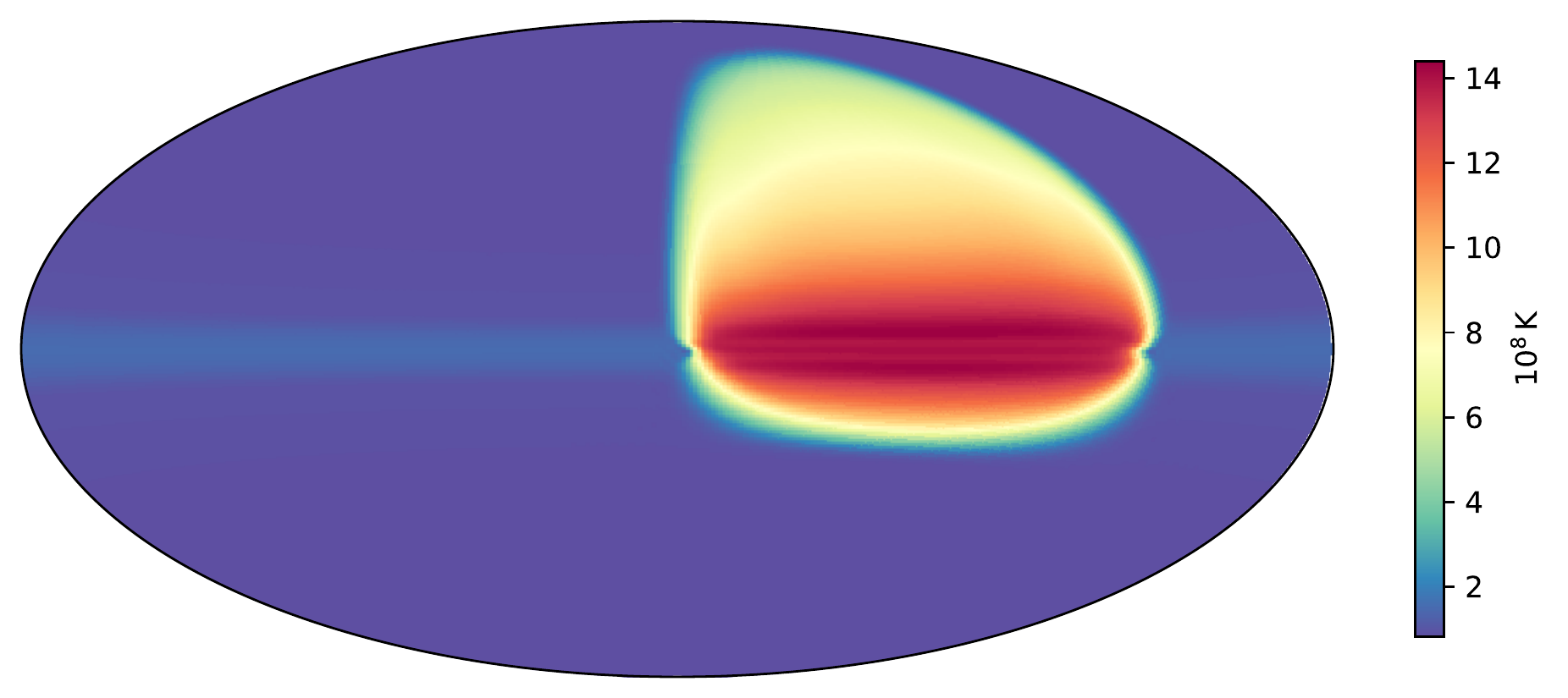}}
\subfigure{\includegraphics[width=.45\textwidth]{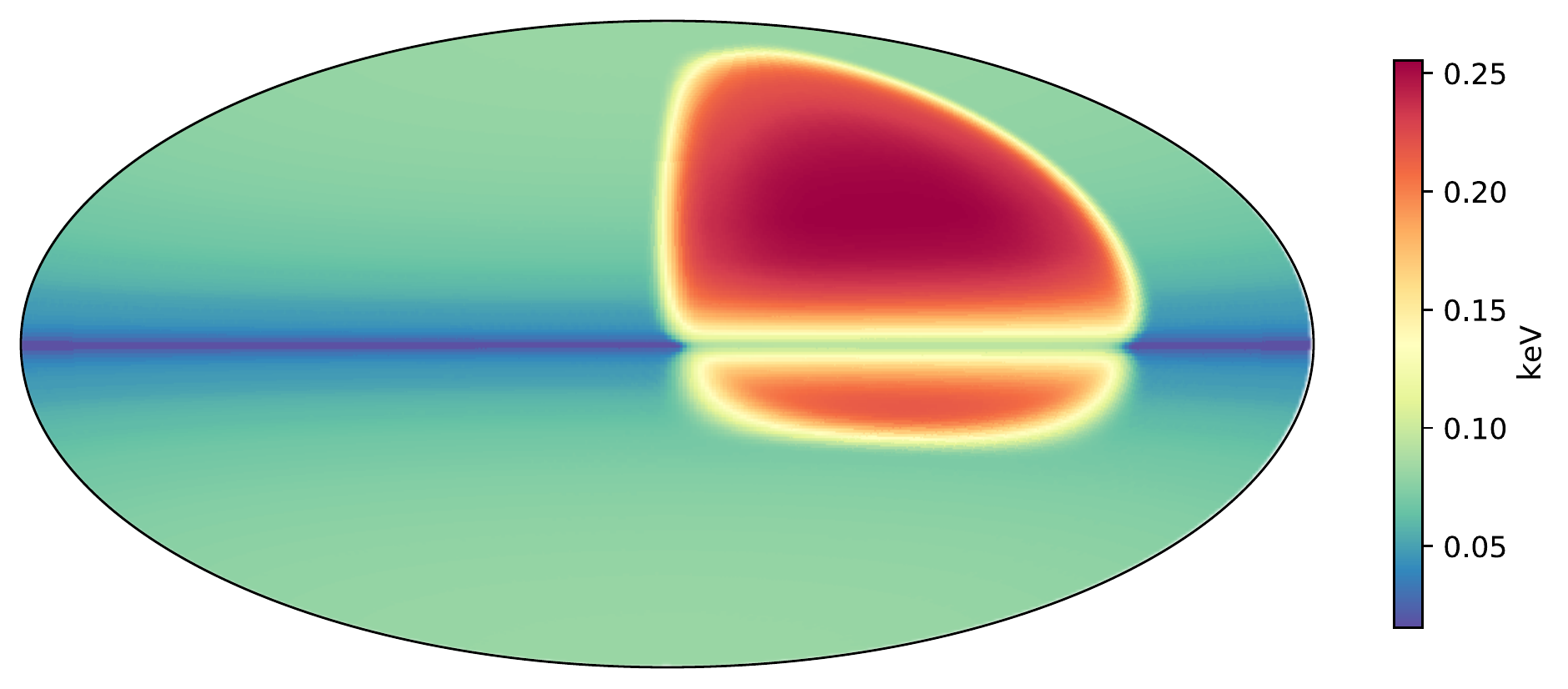}}
\caption{Thermal map at the crust top (upper panel) and at the surface (lower panel) for the second 
lowest curve in Figure\ \ref{fig:saturationCurves}, $H\simeq\SI{1.6e40}{\erg}$, after $t\simeq\SI{9}{\day}$.} \label{fig:32}
\end{figure}

Figure\ \ref{fig:envelope_bursts} shows the unfolding of these outburst models.
The location of the heated zone reflects most substantially on the ensuing evolution. Injecting heat in the equatorial zone results in a longer rise time and overall duration of the event, as well as in a lower peak luminosity. These differences arise because of the thermal insulation properties of the magnetic field. Much in the same way as magnetic insulation in the envelope creates a cold belt where field lines are parallel to the surface \citep[e.g.][]{transport,2020MNRAS.497.2883K}, the diffusion of the injected heat towards the surface is thwarted across field lines inside the crust itself. This causes more heat to be retained in the injection region, which becomes hotter, implying an enhanced neutrino dissipation which, in turn, translates into a lower photon luminosity. Contrariwise, the curves with different envelopes and same location have very similar timing and shape, although with some quantitative differences, albeit on different luminosity scales. 
Actually, not only the energy output, but the very shape of the hot spot changes quite dramatically in the two scenarios. Figure\ \ref{fig:spot_shapes} shows the temperature map during the rise phase both at the top of the crust and at the surface. In the polar case the nearly circular shape of the injection region is preserved, while in the equatorial case the same spot emerges as two separate lobes, that then tend to merge as their luminosity decays. Note, however, that even when a single hot structure is present across the equator it will appear as two spots on the surface due to the efficient magnetic screening in the envelope at the equator itself, where field lines are parallel to the surface.
An intermediate situation is the one investigated in Section \ref{sec:saturation}, with the heated region placed about halfway in the northern hemisphere. The corresponding thermal maps are shown in Figure \ref{fig:32} for the second lowest curve of Figure\ \ref{fig:saturationCurves}, $H\simeq\SI{1.6e40}{\erg}$, after $t\simeq\SI{9}{\day}$. In this case different degrees of heat retention are present in different zones. Even though energy was injected with a maximum at the geometrical center of the patch, the region across the equator retained a larger amount of heat (which again is not directly translated in a hot surface spot due to the effect of the envelope), while the one closer to the pole cooled down more quickly. This results in an asymmetric shape of the surface hot spot, which is likely the most common occurrence in this kind of phenomenon. 

Note that the cases considered here have a quite modest peak luminosity compared to the maximum one allowed before saturation. This has been done on purpose, both in order to reduce at a minimum the necessity to cure neutrino overshooting and to make evident the different behaviors. Once the neutrino saturation regime is entered, in fact, the luminosity curves get so close to each other that the various effects discussed here would be hardly distinguishable.

\subsection{Beyond dipolar fields}\label{sec:WIP}
In the outburst models presented thus far, the background field, assumed to be initially a pure dipole, does not evolve significantly over the short time-span of the events, $\lesssim \SI{1}{\year}$. 
Nevertheless, even with such a simple configuration the key role played by the magnetic topology in the outburst properties became apparent. This naturally prompts for the exploration of different, more complex fields. In this respect, there are at least two aspects to be addressed: a multipolar background field---which may be present in magnetar as the result of magnetic amplification in the proto-NS and/or of the Hall cascade \citep[e.g.][respectively]{2021A&A...645A.109R,2020MNRAS.495.1692G}---and additional field components created by the heated region itself via battery effects, as discussed in \paperI.

\begin{figure}
\centering
\subfigure{\includegraphics[width=.42\textwidth]{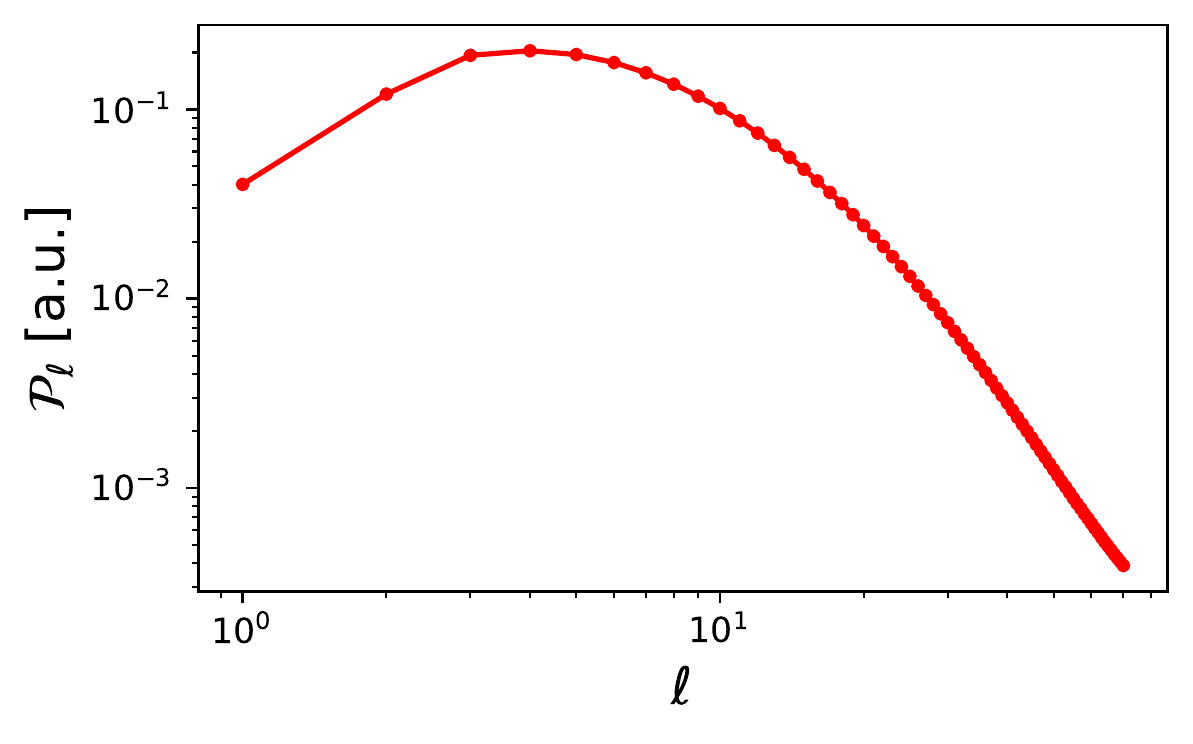}}~\\
\subfigure{{\includegraphics[width=.49\textwidth]{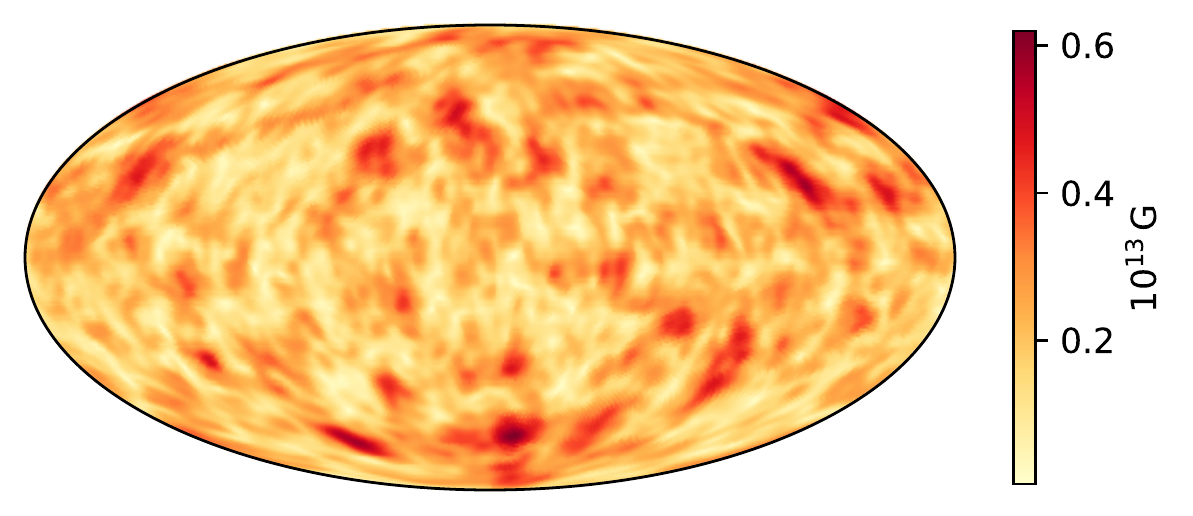}}}
\caption{Power spectrum \emph{(top panel)} and surface strength \emph{(bottom panel)} of a highly multipolar field built with pseudo-random phases.} \label{fig:fancyB}
\end{figure}

In order to explore this scenario, a highly multipolar field was built starting from a given power spectrum $\mathcal{P}_\ell$. To this end, we specified the generic mode of the initial poloidal field as 
\begin{equation}
    B_{\ell m} (r) \sim \frac{W(r)}{\sqrt{\mathcal{P}_{\ell m}}}  \,\cos\left[\frac{2\pi}{\lambda_r} (r-r_c) + \phi_{\ell m}\right] e^{\ii\, \psi_{\ell m}}
\end{equation}
where $\lambda_r=\sqrt{{r^2}/{\ell(\ell+1)}}$ is the typical radial wavelength associated to each mode, $W(r)$ is a weight function that ensures that each component matches the boundary conditions smoothly, $r_c$ is the internal radius of the crust and $\phi_{\ell m}$, $\psi_{\ell m}$ are pseudo-random phases. The desired power spectrum is recovered by modulating each mode with the value $\mathcal{P}_{\ell m}$ itself. The spectrum we used in this case is shown in Figure \ref{fig:fancyB} and was chosen to mimic the one obtained by \citet{2020MNRAS.495.1692G} as the result of the Hall reconfiguration of a field that initially had energy in the small scales only \citep[see also][who obtained spectra with quite similar features as the result of magneto-rotational instability in a proto-NS]{2021A&A...645A.109R}; note in particular that the dipole is not the dominant component.

\begin{figure}
\centering
\subfigure{\includegraphics[width=.47\textwidth]{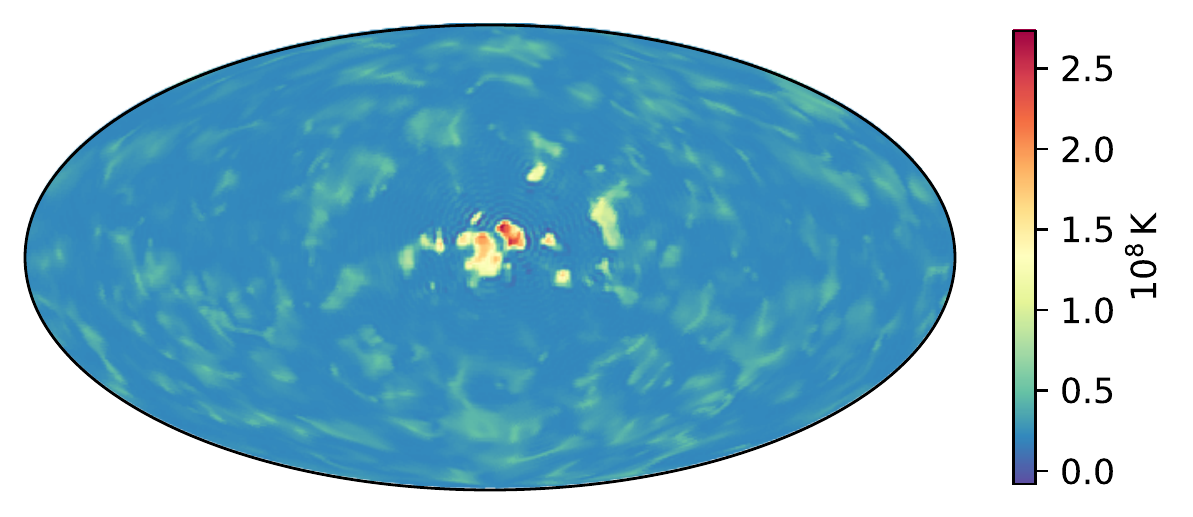}}
\subfigure{\includegraphics[width=.49\textwidth]{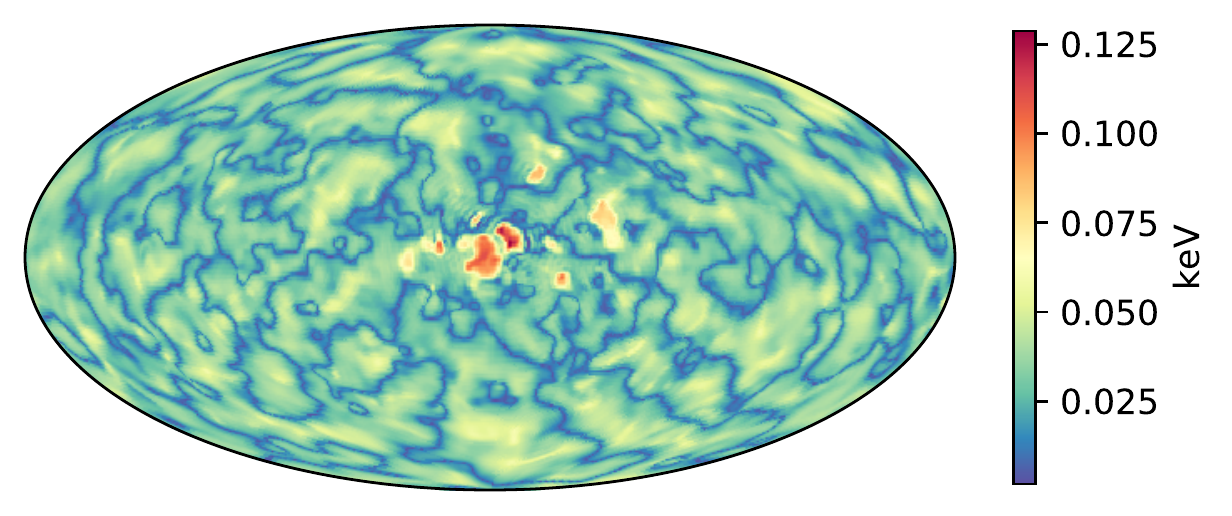}}
\caption{Hot-spot formed after a heat injection event in a model with the multipolar field from Figure\ \ref{fig:fancyB}. The temperature maps are shown in Hammer projection both at the crust top (upper panel) and on the surface (i.e. after applying Eq.\ \ref{eq:tsuruta}, lower panel).} \label{fig:fancySpot}
\end{figure}

A heating source with total input  $H\simeq\SI{5e38}{\erg}$ was placed in a shallow region with $r\simeq\SI{5}{\kilo\meter}$ at $\rho\simeq\SI{6e7}{\gcc}$ across the equator of the coordinate system. Figure \ref{fig:fancySpot} shows the resulting thermal surface map after $\sim\SI{21}{\day}$, i.e.\ during the rising phase of the lightcurve. Although the shape and dimension of the injection are similar to those of the run discussed in the previous section, now the result is the appearance of a number of smaller heated spots on top of a surface distribution which is itself patchy because of the tangled topology of the magnetic field.  
This mirrors, on a different scale, what happens in the dipolar case and is due again to heat flowing towards the surface preferentially where the field lines are mostly radial. Because of the small scale structure, thermal gradients are amplified and, as already mentioned in connection with neutrino losses (see Section \ref{sec:input}), managing large gradients with a spectral code is particularly challenging. For this reason, we could not follow the evolution of this event much beyond the stage shown in Figure \ref{fig:fancySpot}. The results shown here are then preliminary and mostly meant to illustrate a more general magnetic field topology, rather than to provide detailed predictions.
\medskip

\section{Discussion}\label{sec:discussion}
In this work we presented a set of simulations of the magneto-thermal evolution of strongly magnetized neutron stars following an impulsive release of energy in the outer crust. Results were obtained with a fully coupled and 3D approach and are meant to investigate the role played by the many quantities involved in the process, most notably the magnetic field.
We found that the overall behavior observed in magnetar outbursts, e.g.\ the characteristic timescales and the shape of the lightcurves, can be reproduced with a simple model in which heat is deposited over a short timescale ($\lesssim \SI{1}{\hour}$) in a localized region of the outer crust, some kilometers wide and $\approx\SI{100}{\meter}$ thick.

Our results are in substantial agreement with those of \citetalias{2012ApJ...750L...6P}, obtained within a 2D, axially symmetric approach. In particular, we confirm that neutrino losses place an upper limit to the maximum photon luminosity; it is worth noticing that this is not an universal limit but it depends on the details of the injection process. We were also able to verify that the peak luminosity is sensitive to the injection depth and 
heat deposition in the inner crust is unable to produce an outburst-like event, as hinted in \citetalias{2012ApJ...750L...6P}.

Our 3D treatment, nevertheless, goes beyond 
previous studies, in which the heated region had necessary an annular shape, allowing us to investigate how the geometry of the hot spot evolves and how the outburst properties are influenced by the position of the injection zone with respect to the magnetic field. As a general rule, when magnetic field lines run mostly parallel to the surface, heat transport is inefficient and the resulting outburst is weaker (and the onset delayed) with respect to the  case in which the field is more perpendicular to the surface. For the same reason, even in the presence of a simple dipolar field the evolution of the hot spot can be quite different. If heat is injected in the polar region, the hot spot keeps its (nearly circular) shape during the evolution; on the other hand, when heat is supplied at intermediate (magnetic) latitudes, it becomes more complicated and changes in time. More complex field configurations can easily give rise to multiple heated surface regions forming complicated patterns.

In the present work, we focused on the evolution of the local quantities in the NS crust, rather than modeling the ensuing emission as observed by a distant observer. For this reason, the (bolometric) luminosity has been computed by integration over the whole surface of the star, and all the quantities were given in the local frame of reference. However, magnetars spin with periods $P\sim1$--$\SI{10}{\second}$ \citep[e.g.][]{2015RPPh...78k6901T}, so that as the they rotate the visible part of the surface changes. Moreover, light propagating from the surface to the observer is affected by general-relativistic ray bending, so that a proper ray-tracing technique is to be employed to derive the (rotational) phase resolved spectrum as seen at infinity. Such a calculation was carried out in \citetalias{2021ApJ...914..118D} in the case of the long-term evolution of a NS and we plan to do the same for outburst models in an upcoming study.

The heating mechanism responsible for outbursts is commonly believed to be associated to some kind of magnetic energy dissipation, yet it is still unclear what its exact trigger is and, in particular, whether it comes from inside the crust itself. Our simulations show that outbursts can indeed be produced by heating events in the crust and that heat should be released in the outermost, low-density layers. These are precisely the ones more prone to failing due to the build-up of internal magnetic stresses \citep[see e.g.][]{2010MNRAS.407L..54C}. On the other hand, the fact that the top layers of the crust must be involved in order to reproduce the observed timescales may also be compatible with a magnetospheric origin, in which the currents resulting from a rearrangement of the (ultra-strong) magnetic field are heating the crust from above. Nevertheless, our approach is agnostic about the heating mechanism by design, since arbitrarily heat injection we study is intended as a minimal model. Depending on the exact scenario that one may wish to test, further physical ingredients may be added to the model. In particular, the evolution of an outburst, especially in its initial stages, may be affected by short term effects such as shock waves propagating through the crust \citep[e.g.][]{2007ApJ...670.1291W}, heat convection related to a Rayleigh–Taylor instability \citep[e.g.][]{2012ApJ...752..150K} or even perturbations of the magnetic field via MHD disturbances like Alfvén waves; these nevertheless are beyond the scope of this paper and left for future work.

Thanks to their novel 3D approach, our simulations provide a definite step forward in the modeling of outburst physics. Building on this model, there is room for further refinements which will lead to an even more realistic description. In fact, the increased numerical complexity of a 3D setup introduced some trade-offs in the level of microphysical detail which could be incorporated. These concern in particular the treatment of the conductivity, which was based on a constant relaxation time $\tau$, and a schematic treatment of the crustal hydrostatic structure. In particular, we did not include a proper chemical stratification in the outer crust, nor a self-consistent model of the heat blanketing envelope, 
which was rather described via the $T_b$--$T_s$ relation (Eq.\ \ref{eq:tsuruta}) using the numerical fits that are available in the literature. These are computed for thick envelopes rather than the thin one that is implied by the low external density of our models. This results in an underestimate of the surface temperature, as the ensuing thermal gradient is computed across a wider region. Nevertheless, we verified that using envelopes with smaller thicknesses yields qualitatively similar results. The effect of decreasing the density at the base of the envelope is to shift almost rigidly the luminosity curve upwards (by a factor $\approx 5$ in the cases we tested), much in the same way as to what happens when a light element chemical composition is chosen (see Figure \ref{fig:envelope_bursts}). 
Moreover, models of thin envelopes are not available in the literature with magnetic corrections \citep{2016MNRAS.459.1569B}, so that we had to resort to results from \citet{2001PhR...354....1Y} for a thick envelope. In particular, this model tends to underestimate the surface temperature when the field is parallel to the surface (e.g.\ at the equator of a dipolar field). This may artificially  enhance the effect of the splitting of the hot spot across the equator, although we found that the anisotropic conduction within the crust is responsible for this effect by itself, even before the surface is reached (see Figure \ref{fig:spot_shapes}, bottom-left panel).

While this drawback may be overcome by developing a self-consistent magnetized envelope model, or ideally an approach that dismisses the need of a stationary envelope altogether \citep[as e.g.\ the 1D one in][]{2018A&A...609A..74P}, the general picture emerging from our (simplified) model is fairly robust in connection with one of our main goals, i.e.\ the role played by the magnetic field, given also the degeneracy introduced by the many parameters.

Moreover, the hot spots we studied are somewhat larger that what suggested by observations. Smaller heated patches imply steeper thermal gradients for a given total luminosity and this becomes numerically troublesome because of the very nature of spectral schemes, which need extremely high resolutions to resolve sharp gradients. This also hindered the study of the evolution in a highly multipolar field (see Section \ref{sec:WIP}), that naturally produces small hot spots even starting from a relatively large heated region.

A further point to be considered in connection with the appearance of steep thermal gradients is the generation of additional magnetic field components via the Biermann battery effect \citep{1950ZNatA...5...65B}. As already mentioned in Section \ref{sec:input}, the thermopower term, which accounts for the battery effect, was turned off in Equations (\ref{eq:Induction}) and (\ref{eq:heat}). We checked that it is indeed negligible in nearly all the cases we examined, but it starts to play a role for very large injection rates, close to neutrino saturation. This causes the growth of small-scale magnetic structures which quickly drive numerical instabilities. The very presence of these issues suggests that battery fields may be an important feature in magnetar outbursts when the energy input is very large and/or the spot structure is complex. The growth of (strong) small-scale magnetic components may also directly bear to the absorption features observed in the spectra of some magnetars in outburst and interpreted as proton cyclotron lines produced in small magnetic loops \citep[][]{2013Natur.500..312T,10.1093/mnras/stv2490}. Yet, a global model involving the entire crust, like the one at hand, is not the most befitting option to focus on small-scale phenomena. A better understanding of thermopower effects should be gained through dedicated local models, able to efficiently resolve small scales and sharp gradients. 

Despite dealing with a relatively limited set of representative cases, particularly concerning the magnetic field configurations, our results show that the physics behind outbursts is very rich and that their evolution depends on several parameters. Indeed, the known outbursts exhibit rather different characteristics and, in particular, the evolution of their bolometric luminosity changes quite markedly among the events, even those coming from the same source \citep{2018MNRAS.474..961C}. On the other hand, the number of known outbursts to date is relatively small (the Magnetar Outburst Online Catalog, see footnote \ref{note1}, reports $22$ events from $16$ sources) so that a properly data-driven study can not be put into place, especially considering that the parameter space of field geometries and heating conditions is extremely large and cannot be sampled efficiently in the present computational setup, in which a single model takes several hours to compute. The study of the conditions underlying magnetar activity is presently experiencing a renewed interest, thanks in particular to its association to Fast Radio Bursts \citep{2020GCN.27668....1M}; therefore, more and more accurate and efficient models assessing these powerful events will become essential for the understanding of NS astrophysics as a whole.

\begin{acknowledgments}
We thank the anonymous Referee for several insightful comments which contributed to improve our manuscript. Simulations were run at CloudVeneto, a HPC facility jointly owned by the University of Padova and INFN, and at UCL Myriad HPC facility (Myriad@UCL). The authors gratefully acknowledge the use of both facilities and the associated support services. RT and RT acknowledge financial support from the Italian MUR through grant PRIN 2017LJ39LM.
\end{acknowledgments}

%

\bibliography{biblio.bib}{}

\begin{thebibliography}{}
\expandafter\ifx\csname natexlab\endcsname\relax\def\natexlab#1{#1}\fi
\providecommand{\url}[1]{\href{#1}{#1}}
\providecommand{\dodoi}[1]{doi:~\href{http://doi.org/#1}{\nolinkurl{#1}}}
\providecommand{\doeprint}[1]{\href{http://ascl.net/#1}{\nolinkurl{http://ascl.net/#1}}}
\providecommand{\doarXiv}[1]{\href{https://arxiv.org/abs/#1}{\nolinkurl{https://arxiv.org/abs/#1}}}

\bibitem[{{Beloborodov}(2009)}]{2009ApJ...703.1044B}
{Beloborodov}, A.~M. 2009, \apj, 703, 1044,
  \dodoi{10.1088/0004-637X/703/1/1044}

\bibitem[{{Beznogov} {et~al.}(2016){Beznogov}, {Potekhin}, \&
  {Yakovlev}}]{2016MNRAS.459.1569B}
{Beznogov}, M.~V., {Potekhin}, A.~Y., \& {Yakovlev}, D.~G. 2016, \mnras, 459,
  1569, \dodoi{10.1093/mnras/stw751}

\bibitem[{{Beznogov} {et~al.}(2021){Beznogov}, {Potekhin}, \&
  {Yakovlev}}]{2021PhR...919....1B}
---. 2021, \physrep, 919, 1, \dodoi{10.1016/j.physrep.2021.03.004}

\bibitem[{{Biermann}(1950)}]{1950ZNatA...5...65B}
{Biermann}, L. 1950, Zeitschrift Naturforschung Teil A, 5, 65

\bibitem[{{Chugunov} \& {Horowitz}(2010)}]{2010MNRAS.407L..54C}
{Chugunov}, A.~I., \& {Horowitz}, C.~J. 2010, \mnras, 407, L54,
  \dodoi{10.1111/j.1745-3933.2010.00903.x}

\bibitem[{{Coti Zelati} {et~al.}(2018){Coti Zelati}, {Rea}, {Pons}, {Campana},
  \& {Esposito}}]{2018MNRAS.474..961C}
{Coti Zelati}, F., {Rea}, N., {Pons}, J.~A., {Campana}, S., \& {Esposito}, P.
  2018, \mnras, 474, 961, \dodoi{10.1093/mnras/stx2679}

\bibitem[{{De Grandis} {et~al.}(2021){De Grandis}, {Taverna}, {Turolla},
  {Gnarini}, {Popov}, {Zane}, \& {Wood}}]{2021ApJ...914..118D}
{De Grandis}, D., {Taverna}, R., {Turolla}, R., {et~al.} 2021, \apj, 914, 118,
  \dodoi{10.3847/1538-4357/abfdac}

\bibitem[{{De Grandis} {et~al.}(2020){De Grandis}, {Turolla}, {Wood}, {Zane},
  {Taverna}, \& {Gourgouliatos}}]{2020ApJ...903...40D}
{De Grandis}, D., {Turolla}, R., {Wood}, T.~S., {et~al.} 2020, \apj, 903, 40,
  \dodoi{10.3847/1538-4357/abb6f9}

\bibitem[{Dormy(1997)}]{dormy}
Dormy, E. 1997, PhD thesis, Institut de Physique du Globe de Paris.
\newblock \url{http://www.math.ens.fr/~dormy/these/index.html}

\bibitem[{{Esposito} {et~al.}(2021){Esposito}, {Rea}, \&
  {Israel}}]{2021ASSL..461...97E}
{Esposito}, P., {Rea}, N., \& {Israel}, G.~L. 2021, in Astrophysics and Space
  Science Library, Vol. 461, Astrophysics and Space Science Library, ed. T.~M.
  {Belloni}, M.~{M{\'e}ndez}, \& C.~{Zhang}, 97--142,
  \dodoi{10.1007/978-3-662-62110-3\_3}

\bibitem[{{Gourgouliatos} \& {Cumming}(2014)}]{2014MNRAS.438.1618G}
{Gourgouliatos}, K.~N., \& {Cumming}, A. 2014, \mnras, 438, 1618,
  \dodoi{10.1093/mnras/stt2300}

\bibitem[{{Gourgouliatos} {et~al.}(2020){Gourgouliatos}, {Hollerbach}, \&
  {Igoshev}}]{2020MNRAS.495.1692G}
{Gourgouliatos}, K.~N., {Hollerbach}, R., \& {Igoshev}, A.~P. 2020, \mnras,
  495, 1692, \dodoi{10.1093/mnras/staa1295}

\bibitem[{{Gourgouliatos} \& {Lander}(2021)}]{2021MNRAS.506.3578G}
{Gourgouliatos}, K.~N., \& {Lander}, S.~K. 2021, \mnras, 506, 3578,
  \dodoi{10.1093/mnras/stab1869}

\bibitem[{{Gudmundsson} {et~al.}(1983){Gudmundsson}, {Pethick}, \&
  {Epstein}}]{1983ApJ...272..286G}
{Gudmundsson}, E.~H., {Pethick}, C.~J., \& {Epstein}, R.~I. 1983, \apj, 272,
  286, \dodoi{10.1086/161292}

\bibitem[{{Igoshev} {et~al.}(2020){Igoshev}, {Hollerbach}, {Wood}, \&
  {Gourgouliatos}}]{2020NatAs.tmp..215I}
{Igoshev}, A.~P., {Hollerbach}, R., {Wood}, T., \& {Gourgouliatos}, K.~N. 2020,
  Nature Astronomy, \dodoi{10.1038/s41550-020-01220-z}

\bibitem[{Karageorgopoulos {et~al.}(2019)Karageorgopoulos, Gourgouliatos, \&
  Contopoulos}]{10.1093/mnras/stz1507}
Karageorgopoulos, V., Gourgouliatos, K.~N., \& Contopoulos, I. 2019, \mnras,
  487, 3333, \dodoi{10.1093/mnras/stz1507}

\bibitem[{{Kaspi} \& {Beloborodov}(2017)}]{2017ARA&A..55..261K}
{Kaspi}, V.~M., \& {Beloborodov}, A.~M. 2017, \araa, 55, 261,
  \dodoi{10.1146/annurev-astro-081915-023329}

\bibitem[{{Keek} {et~al.}(2012){Keek}, {Heger}, \& {in't
  Zand}}]{2012ApJ...752..150K}
{Keek}, L., {Heger}, A., \& {in't Zand}, J.~J.~M. 2012, \apj, 752, 150,
  \dodoi{10.1088/0004-637X/752/2/150}

\bibitem[{{Kondratyev} {et~al.}(2020){Kondratyev}, {Moiseenko},
  {Bisnovatyi-Kogan}, \& {Glushikhina}}]{2020MNRAS.497.2883K}
{Kondratyev}, I.~A., {Moiseenko}, S.~G., {Bisnovatyi-Kogan}, G.~S., \&
  {Glushikhina}, M.~V. 2020, \mnras, 497, 2883, \dodoi{10.1093/mnras/staa2154}

\bibitem[{{Lander}(2016)}]{2016ApJ...824L..21L}
{Lander}, S.~K. 2016, \apjl, 824, L21, \dodoi{10.3847/2041-8205/824/2/L21}

\bibitem[{{Lander} \& {Gourgouliatos}(2019)}]{2019MNRAS.486.4130L}
{Lander}, S.~K., \& {Gourgouliatos}, K.~N. 2019, \mnras, 486, 4130,
  \dodoi{10.1093/mnras/stz1042}

\bibitem[{{Mereghetti} {et~al.}(2020){Mereghetti}, {Savchenko}, {Gotz},
  {Ducci}, {Ferrigno}, {Bozzo}, {Borkowski}, \&
  {Bazzano}}]{2020GCN.27668....1M}
{Mereghetti}, S., {Savchenko}, V., {Gotz}, D., {et~al.} 2020, GRB Coordinates
  Network, 27668, 1

\bibitem[{{Perna} \& {Pons}(2011)}]{2011ApJ...727L..51P}
{Perna}, R., \& {Pons}, J.~A. 2011, \apjl, 727, L51,
  \dodoi{10.1088/2041-8205/727/2/L51}

\bibitem[{{Pons} {et~al.}(2009){Pons}, {Miralles}, \&
  {Geppert}}]{2009A&A...496..207P}
{Pons}, J.~A., {Miralles}, J.~A., \& {Geppert}, U. 2009, \aap, 496, 207,
  \dodoi{10.1051/0004-6361:200811229}

\bibitem[{{Pons} \& {Rea}(2012)}]{2012ApJ...750L...6P}
{Pons}, J.~A., \& {Rea}, N. 2012, \apjl, 750, L6,
  \dodoi{10.1088/2041-8205/750/1/L6}

\bibitem[{{Potekhin} \& {Chabrier}(2018)}]{2018A&A...609A..74P}
{Potekhin}, A.~Y., \& {Chabrier}, G. 2018, \aap, 609, A74,
  \dodoi{10.1051/0004-6361/201731866}

\bibitem[{{Potekhin} {et~al.}(2015){Potekhin}, {Pons}, \& {Page}}]{transport}
{Potekhin}, A.~Y., {Pons}, J.~A., \& {Page}, D. 2015, Space Science Reviews,
  191, 239

\bibitem[{{Potekhin} \& {Yakovlev}(2001)}]{envelopes}
{Potekhin}, A.~Y., \& {Yakovlev}, D.~G. 2001, A\&A, 374, 213,
  \dodoi{10.1051/0004-6361:20010698}

\bibitem[{{Reboul-Salze} {et~al.}(2021){Reboul-Salze}, {Guilet}, {Raynaud}, \&
  {Bugli}}]{2021A&A...645A.109R}
{Reboul-Salze}, A., {Guilet}, J., {Raynaud}, R., \& {Bugli}, M. 2021, \aap,
  645, A109, \dodoi{10.1051/0004-6361/202038369}

\bibitem[{Rodríguez~Castillo {et~al.}(2016)Rodríguez~Castillo, Israel,
  Tiengo, Salvetti, Turolla, Zane, Rea, Esposito, Mereghetti, Perna, Stella,
  Pons, Campana, Götz, \& Motta}]{10.1093/mnras/stv2490}
Rodríguez~Castillo, G.~A., Israel, G.~L., Tiengo, A., {et~al.} 2016, Monthly
  Notices of the Royal Astronomical Society, 456, 4145,
  \dodoi{10.1093/mnras/stv2490}

\bibitem[{{Thompson}(2001)}]{2001nsbh.conf..369T}
{Thompson}, C. 2001, in The Neutron Star - Black Hole Connection, ed.
  C.~{Kouveliotou}, J.~{Ventura}, \& E.~{van den Heuvel}, Vol. 567, 369.
\newblock \doarXiv{astro-ph/0010016}

\bibitem[{{Tiengo} {et~al.}(2013){Tiengo}, {Esposito}, {Mereghetti}, {Turolla},
  {Nobili}, {Gastaldello}, {G{\"o}tz}, {Israel}, {Rea}, {Stella}, {Zane}, \&
  {Bignami}}]{2013Natur.500..312T}
{Tiengo}, A., {Esposito}, P., {Mereghetti}, S., {et~al.} 2013, \nat, 500, 312,
  \dodoi{10.1038/nature12386}

\bibitem[{{Turolla} {et~al.}(2015){Turolla}, {Zane}, \&
  {Watts}}]{2015RPPh...78k6901T}
{Turolla}, R., {Zane}, S., \& {Watts}, A.~L. 2015, Reports on Progress in
  Physics, 78, 116901, \dodoi{10.1088/0034-4885/78/11/116901}

\bibitem[{{Weinberg} \& {Bildsten}(2007)}]{2007ApJ...670.1291W}
{Weinberg}, N.~N., \& {Bildsten}, L. 2007, \apj, 670, 1291,
  \dodoi{10.1086/522111}

\bibitem[{{Wood} \& {Hollerbach}(2015)}]{2015PhRvL.114s1101W}
{Wood}, T.~S., \& {Hollerbach}, R. 2015, \prl, 114, 191101,
  \dodoi{10.1103/PhysRevLett.114.191101}

\bibitem[{{Yakovlev} {et~al.}(2001){Yakovlev}, {Kaminker}, {Gnedin}, \&
  {Haensel}}]{2001PhR...354....1Y}
{Yakovlev}, D.~G., {Kaminker}, A.~D., {Gnedin}, O.~Y., \& {Haensel}, P. 2001,
  \physrep, 354, 1, \dodoi{10.1016/S0370-1573(00)00131-9}

\bibitem[{{Yakovlev} {et~al.}(2021){Yakovlev}, {Kaminker}, {Potekhin}, \&
  {Haensel}}]{2021MNRAS.500.4491Y}
{Yakovlev}, D.~G., {Kaminker}, A.~D., {Potekhin}, A.~Y., \& {Haensel}, P. 2021,
  \mnras, 500, 4491, \dodoi{10.1093/mnras/staa3547}

\end{thebibliography}
\bibliographystyle{aasjournal}



\end{document}